\documentclass{article}

\usepackage{PRIMEarxiv}

\usepackage[utf8]{inputenc} 
\usepackage[T1]{fontenc}    
\usepackage{hyperref}       
\usepackage{url}            
\usepackage{booktabs}       
\usepackage{amsfonts}       
\usepackage{nicefrac}       
\usepackage{microtype}      
\usepackage{lipsum}
\usepackage{fancyhdr}       
\usepackage{graphicx}       
\graphicspath{{media/}}     

\usepackage{lmodern}
\usepackage{amsmath}

\title{Reduced-order modeling of a viscoelastic turbulent jet with hybrid machine learning models
}

\author{
  Christian Amor\\
  Complex Fluids and Flows Unit \\
  Okinawa Institute of Science and Technology Graduate University \\
  Okinawa, Japan\\
  \And
  Adri{á}n Corrochano \\
  School of Aerospace Engineering \\
  Universidad Polit{é}cnica de Madrid \\
  Madrid, Spain\\
  \And
  Marco Edoardo Rosti\\
  Complex Fluids and Flows Unit \\
  Okinawa Institute of Science and Technology Graduate University \\
  Okinawa, Japan\\
  \texttt{marco.rosti@oist.jp}
  \And
  Soledad Le Clainche \\
  School of Aerospace Engineering \\
  Universidad Polit{é}cnica de Madrid \\
  Madrid, Spain\\
  \texttt{soledad.leclainche@upm.es}
}

\begin{document}
\maketitle

\begin{abstract}
Adding flexible polymers to a Newtonian solvent confers complex properties to the resulting solution. The additional complexity substantially increases the computational cost of numerical simulations, which often makes them prohibitively expensive. Here, we propose hybrid reduced-order models to accelerate simulations of viscoelastic turbulent jets. The model combines modal decompositions with deep networks: we use proper orthogonal decomposition to obtain a compact representation of the data, and a neural network is trained to predict the mode coefficients in the low-dimensional space. Results show that the hybrid model effectively captures the long-term behavior of the viscoelastic jet, that we demonstrate by computing relevant statistics of the jet. While small models are capable of predicting large-scale dynamics more than one-step at a time, thus facilitating greater accelerations, larger models are mandatory for forecasting smaller-scale dynamics, with skip connections the most effective strategy for deeper and generalizable models. The proposed methodology underpins the potential of hybrid approaches for compact and robust reduced-order models of viscoelastic turbulent jets.
\end{abstract}

\section{Introduction}
Viscoelastic fluids have an extensive presence in nature and engineering applications. Their unique dynamics makes them technologically interesting, where even a small concentration of polymers can significantly change the properties of Newtonian flows. For instance, polymers can induce drag reduction at high Reynolds numbers $Re$ \cite{Serafini2022PRLTDR}, have complex interactions with inertial dynamics \cite{rosti2023large}, or even destabilize laminar flows through elastic instabilities \cite{Larson1992RheoActaElInst}, that can lead to elastic turbulence at low $Re$ and high polymer elasticity \cite{Groisman2000NatureET, Singh2024NatCommET}. In jet flows, polymers reduce the spreading rate and centerline velocity decay of high-$Re$ viscoelastic jets \cite{guimaraes2020direct}, while at lower $Re$ elastic instabilities trigger a disordered flow that exhibits structures and dynamics significantly different from Newtonian jets \cite{yamani2021spectral, yamani2023spatiotemporal, soligo2023non}. 

Understanding the behavior of polymer-induced turbulence is challenging due to their complex spatio-temporal dynamics, even more so in elastic turbulence. Experimental realizations of elastic turbulence are well documented \cite{Steinberg2021ARFMET}, although measurements of elastic stresses are difficult due to instrumental limitations. In this case, direct numerical simulations can provide full access to the velocity, pressure, and elastic stress fields. However, large-scale simulations are complicated at high Weissenberg numbers $Wi$---ratio of elastic to viscous forces---where numerical instabilities make simulations struggle for $Wi > 1$ \cite{keunings1986high}. Overcoming these instabilities requires sophisticated numerical methods that increase exponentially the cost of numerical simulations, which added to the small time step needed at very low $Re$ often makes them prohibitively expensive.

In this scenario, machine learning can help reducing the computational cost. In particular, reduced-order models (ROMs) can capture the key underlying dynamics of the flow while maintaining an acceptable level of accuracy \cite{Brunton2020ARFMML4FM}. A common approach for ROMs of turbulent flows relies on modal decompositions, whose optimal linear basis is well-suited for surrogate models \cite{Taira2017AAIAJModalAna}. More recent ROMs implement modern machine learning techniques, where purely machine learning ROMs can compress more efficiently turbulent data exploiting nonlinear low-dimensional basis \cite{Murata2020JFMConvAE, Eivazi2022ESWAbVAE}. Then, the temporal dynamics of the nonlinear latent spaces can be modeled using neural networks \cite{Eivazi2020PoFNonLinROM, SoleraRico2024NatCommubVAETrans}. 

However, training purely machine learning ROMs requires big data, that increase with the size of the model. A valid approach for reducing the amount of necessary data considers training the model with additional information obtained, for example, by enforcing physical laws. These can be imposed from the data or through soft penalty constraints that tell the model to adhere to the underlying physics of the system. As a result, embedding physics into purely machine learning ROMs can reduce training time and improve the generalizability of the models \cite{Karniadakis2021NatRevPhysPIML}.

Here, we follow these principles, where we embed physics into our machine learning model through modal decompositions for reduced-order modeling of a viscoelastic turbulent jet. Few past examples of ROMs of viscoelastic flows also considered modal decompositions: they used proper orthogonal decomposition (POD) \cite{Lumley1970POD}, where they exploited the linear subspace for Galerkin methods \cite{Wang2020ICHMTPODGarVET}, built a model using a library of candidate functions for predicting the mode coefficients \cite{Oishi2024RSOPSINDyVET}, or coupled POD with an autoencoder, thus predicting the dynamics of the nonlinear subspace \cite{Kumar2025JFMEITROM}. In this work, we also couple POD with a deep network, though we predict the mode coefficients from POD. In doing so, we drastically reduce the size of the model and consequently the training time and the amount of training data. The optimal subspace from POD is physically-interpretable, and using it during training constitutes a method for weakly enforcing the physics underlying in the system into the machine learning model. We demonstrate that this approach is sufficient for predicting the large-scale dynamics of the turbulent viscoelastic jets accurately, leading to compact and robust ROMs for viscoelastic flows. This work is thus a first attempt of building ROMs for elastic turbulence in three-dimensional viscoelastic jets.

The paper is organized as follows. Section \ref{sec:meth} introduces the methodology used in this work; in particular, we describe the dataset, the numerical simulation, and the prediction models. Section \ref{sec:res} shows the obtained results, and \S\ref{sec:concl} summarizes the main findings.

\section{Methods}\label{sec:meth}
\subsection{Direct numerical simulations}
The incompressible, three-dimensional planar jet of a polymer solution is simulated by means of direct numerical simulations using our in-house solver \textit{Fujin} (\url{https://www.oist.jp/research/research-units/cffu/fujin}). The incompressible Navier-Stokes equations are solved on a staggered, uniform, Cartesian grid, with \textit{x} the streamwise, \textit{y} the jet-normal, and \textit{z} the spanwise directions. The momentum equation is corrected by adding the divergence of the non-Newtonian stress tensor $\tau$, modeled by the Oldroyd-B model: 
\begin{equation}
    \lambda \left( \partial_t \tau + \boldsymbol{u} \cdot \nabla \tau - \nabla \boldsymbol{u}^{\top} \cdot \tau - \tau \cdot \nabla \boldsymbol{u} \right) + \tau = \mu_p \left( \nabla \boldsymbol{u} + \nabla \boldsymbol{u}^{\top} \right),
    \label{eq:a1}
\end{equation}
where $\lambda$ is the relaxation time of the polymer---the time required by the polymer to relax back to equilibrium after being perturbed by an external forcing---and $\mu_p$ the dynamic viscosity of the polymer. Note that eq.~(\ref{eq:a1}) only considers elastic effects in the fluid, thus neglecting shear-dependent viscosity. The non-Newtonian stress $\tau$ is rewritten in terms of the conformation tensor $\mathbf{C}$, a second-order, positive-definite tensor that indicates the average value of the end-to-end distance of the polymers: $\tau = \mu_p \left( \mathbf{C} - \mathbf{I} \right) / \lambda$, with $\mathbf{I}$ the tensorial identity.

Equations are discretized in space using the second-order, central finite differences, and they are advanced in time using a second-order explicit Adams-Bashforth scheme, that is coupled with a fractional step method \cite{kim1985application} to enforce the incompressibility. The transport equation for the polymer conformation tensor $\mathbf{C}$ is carefully solved using a matrix-logarithm formulation \cite{fattal2004constitutive, hulsen2005flow} coupled with a high-order weighted essentially non-oscillatory scheme \cite{shu2009high, sugiyama2011full} for the upper-convective derivative in the left-hand side of eq.~(\ref{eq:a1}) to deal with the high-$Wi$ problem mentioned before.

The computational domain and boundary conditions are as follows. The fluid is injected through a plane slit of height $h$ in a domain with size $160h \times 240h \times 13.33h$, that is discretized using $1440 \times 2340 \times 128$ grid points. The inlet boundary has no-slip and no-penetration conditions, except for the inlet portion, where it is imposed a plug flow with constant velocity $U$, while the outlet boundary has a non-reflective outflow condition \cite{orlanski1976simple}. The upper and lower boundaries ($y = 0$, $y = L_y$) have free-slip and no-penetration conditions, and periodicity is imposed in the lateral boundaries ($z = 0$, $z = L_z$). 

The viscoelastic jet is characterized by three non-dimensional parameters. The first one is the Reynolds number, which is based on conditions at the inlet and is set equal to $Re = U h / \nu_0 = 20$, with $\nu_0$ the total kinematic viscosity of the fluid. The second one is the Weissenberg number, that is fixed to $Wi = U \lambda / h = 100$. The ratio between $Wi$ and $Re$ is greater than one, thus indicating that the flow is dominated by elastic effects rather than inertia. The last parameter is the ratio of the solvent to total viscosity, which is equal to $\beta = \mu_s / \mu_0 = 0.98$, indicating that the polymer solution is dilute.

Full details of the numerical simulation and dataset are described in ref.~\cite{soligo2023non}.

\subsection{Proper orthogonal decomposition} \label{subsec:pod}
We employ POD \cite{Lumley1970POD} to reduce the dimensionality of the data and, in particular, we use the method of snapshots for computing the POD modes \cite{Sirovich1987SVD}. POD finds the subspace that optimally describes the data in the sense that the mean-squared difference between the data and their projection onto the subspace of POD modes is minimized. 

Given a data set of $K$ snapshots, $\mathbf{x}_1, \ldots, \mathbf{x}_K$, we first sample the three-dimensional velocity field from the full domain, where each snapshot contains the streamwise $\boldsymbol{u} \left( \boldsymbol{x}, t \right)$, jet-normal $\boldsymbol{v} \left( \boldsymbol{x}, t \right)$, and spanwise $\boldsymbol{w} \left( \boldsymbol{x}, t \right)$ velocity components. Each component is decomposed as
\begin{equation}
    \boldsymbol{u} \left( \boldsymbol{x}, t \right) = \bar{\boldsymbol{u}} \left( \boldsymbol{x} \right) + \boldsymbol{u}^{\prime} \left( \boldsymbol{x}, t \right), \quad
    \boldsymbol{v} \left( \boldsymbol{x}, t \right) = \bar{\boldsymbol{v}} \left( \boldsymbol{x} \right) + \boldsymbol{v}^{\prime} \left( \boldsymbol{x}, t \right), \quad
    \boldsymbol{w} \left( \boldsymbol{x}, t \right) = \bar{\boldsymbol{w}} \left( \boldsymbol{x} \right) + \boldsymbol{w}^{\prime} \left( \boldsymbol{x}, t \right), 
    \label{eq:b1}
\end{equation}
with $\bar{\boldsymbol{u}} \left( \boldsymbol{x} \right)$, $\bar{\boldsymbol{v}} \left( \boldsymbol{x} \right)$ and $\bar{\boldsymbol{w}} \left( \boldsymbol{x} \right)$ being the time-averaged velocity fields and $\boldsymbol{u}^{\prime} \left( \boldsymbol{x}, t \right)$, $\boldsymbol{v}^{\prime} \left( \boldsymbol{x}, t \right)$ and $\boldsymbol{w}^{\prime} \left( \boldsymbol{x}, t \right)$ the fluctuations, respectively. The fluctuating quantities are concatenated for each snapshot and organized in a matrix $\mathbf{X}$ whose columns are the snapshots $\mathbf{x}_k$. The (truncated) singular value decomposition (SVD) of $\mathbf{X}$ is given by:  
\begin{equation}
    \mathbf{X} \simeq \mathbf{U} \mathbf{\Sigma} \mathbf{V}^{\top} = 
    \sum_{n = 1}^N \sigma_n \mathbf{u}_n \mathbf{v}_n^{\top},
    \label{eq:b2}
\end{equation}
where $\mathbf{\Sigma}$ is a diagonal matrix with entries $\sigma_n > 0$ that are optimally ranked, $\sigma_1 \geq \ldots \geq \sigma_N$, while the columns of $\mathbf{U}$ and $\mathbf{V}$ are orthonormal, so they satisfy the condition $\mathbf{U}^{\top} \mathbf{U} = \mathbf{V}^{\top} \mathbf{V} = \mathbf{I}$. The eigenvalues $\sigma_n$ indicate the energy content of each POD mode, while $\mathbf{u}_n$, and $\mathbf{v}_n$ contain the information of the temporal evolution. The reconstruction error from SVD is reduced for larger $N$, where $N \leq r$, with $r$ the rank of the snapshot matrix.

\subsection{Time series prediction model}
\begin{figure}[t!]
    \centering
    \includegraphics[width=\textwidth, keepaspectratio]{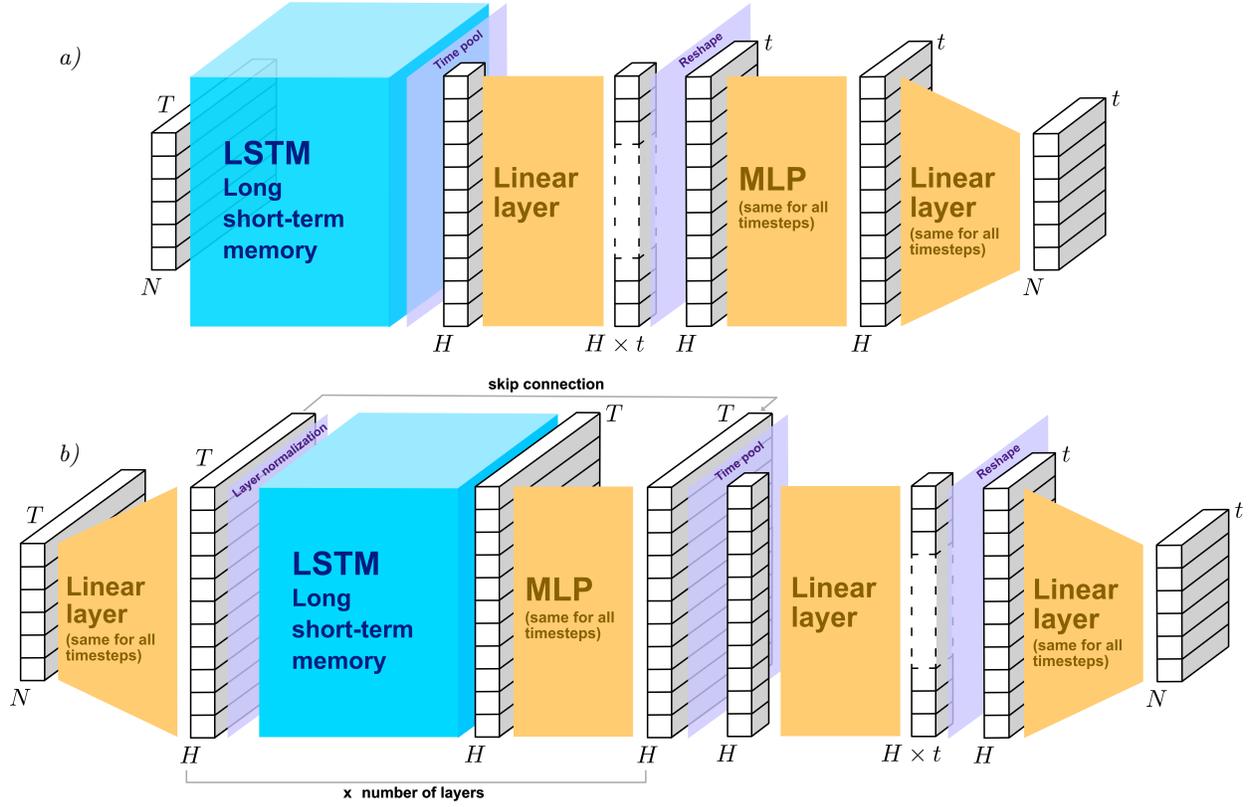}
    \put(-450, 280){\small \textit{a)}}
    \put(-450, 130){\small \textit{b)}}
    \caption{\textbf{Sketch of the prediction models:} (\textit{a}) POD-DL and (\textit{b}) POD-rDL. The dimension of the output for each layer is indicated in each block. The symbols $N$ and $H$ denote the dimensionality, and $T$ and $t$ the length of the sequence.} 
    \label{fig:models}
\end{figure}
Figure~\ref{fig:models} shows a sketch of the prediction models used in this work. The models are trained to use a patch of data to forecast the next time step, whereas predictions are done in an autoregressive way during inference. The input length is set to $T = 64$ and the output is the next time step, though the models can be tuned to predict $t > 1$ steps at a time, hence longer output sequences. In doing this, they go through fewer autoregressive generations given the same task, that results in greater accelerations. Both models are optimized during training for minimizing the difference between the predicted and the true original data, computed based on the mean-squared-error loss function.

We consider two models in this work, the first one being based on the model introduced in ref.~\cite{AbadiaHeredia2022ESWAPODDL}, namely POD-DL (see panel \textit{a}), which combines recurrent and fully-connected layers. At first, a long short-term memory (LSTM) network learns the long-range temporal structure of the input (LSTM has been shown to be effective for predicting turbulent flows \cite{Srinivasan2019PRFLSTM, Nakamura2021POFLSTM}). Then, the output of the LSTM block passes through a multilayer perceptron (MLP)---a stack of fully-connected layers---that learns nonlinear relations between the input and output sequences. The MLP consists of two linear transformations with a ReLU activation in between; the linear transformations are the same across different timesteps. The dimensionality of the input and output is $H = 256$ (same as the LSTM) and the inner-layer has dimensionality $d_{mlp} = 1024$. 

We also explore the effect of layer depth on the predictions in two ways: we either stack several layers within the LSTM block or resort to skip (or residual) connections \cite{He2016Residual}. In the latter case, we introduce a new model termed POD-DL with residual, or POD-rDL for simplicity (see panel \textit{b}), where a skip connection is enabled around each block of LSTM and MLP. In doing this, the network learns how to change the input features rather than overwriting them. As a result, gradients can flow easily through the skip connection, helping stability and preserving useful information across depth. To facilitate these, all sub-layers within the residual block produce outputs of dimension $H$. We also employ pre-layer normalization \cite{Ba2016arxivLayerNorm}, with the output of the residual block given by $x + f( {\rm LayerNorm} (x))$, with $x$ being the input. In the case of the POD-DL, we apply a dropout to the output of each sub-layer within the LSTM block if the number of layers is greater than one, with a rate $P_{drop} = 0.3$ to avoid overfitting. On the contrary, we configure the POD-rDL with a single-layer LSTM, thus not requiring dropout; in this case, layer normalization and skip connections guarantee enough regularization in the model.

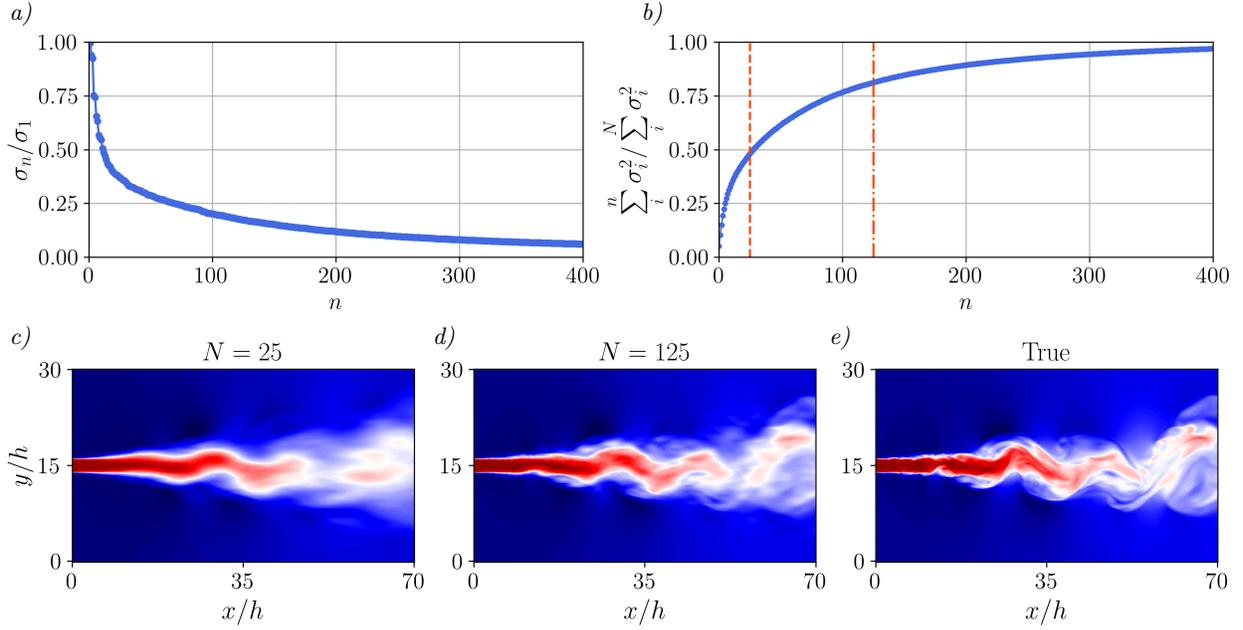
\begin{figure}[t!]
    \centering
    \includegraphics[width=\textwidth, keepaspectratio]{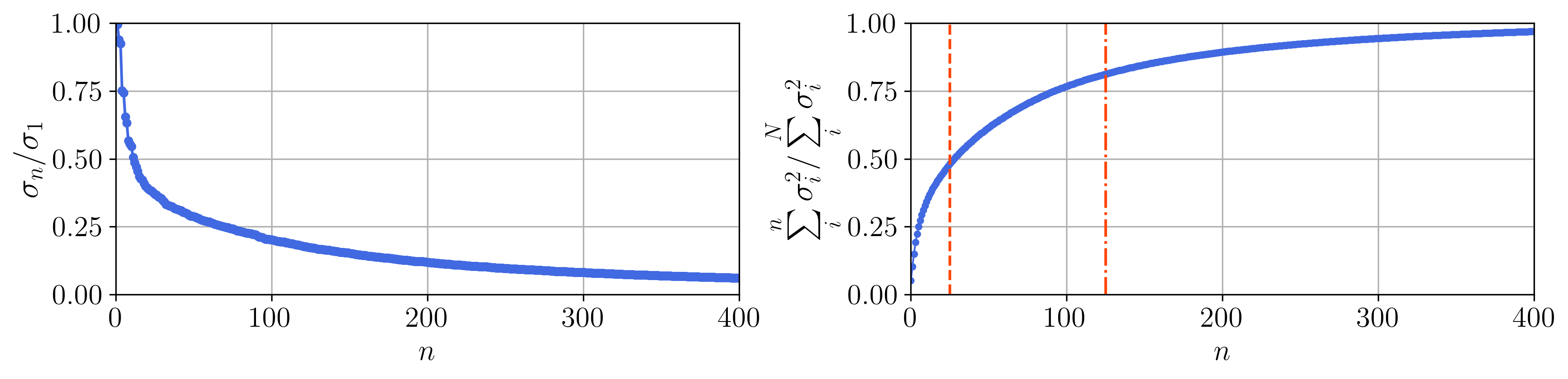}
    \put(-465,115){\small \textit{a)}}
    \put(-226,115){\small \textit{b)}}

    \includegraphics[width=\textwidth, keepaspectratio]{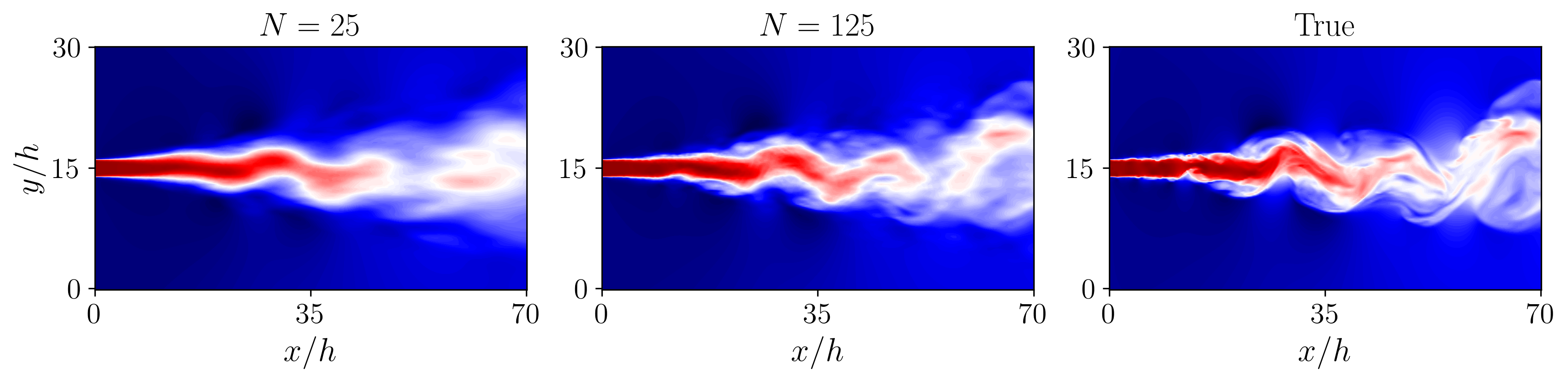}
    \put(-465,110){\small \textit{c)}}
    \put(-305,110){\small \textit{d)}}
    \put(-155,110){\small \textit{e)}}
    \caption{\textbf{POD of the viscoelastic jet}. (\textit{a}) Mode decay and (\textit{b}) fraction of energy as a function of the number of modes. The reconstructions of the streamwise velocity field with (\textit{c}) $25$ and (\textit{d}) $125$ POD modes are compared with ($\textit{e}$) the original flow, with each reconstruction containing $\approx 50\%$ (\textit{b}, dashed red line) and $\approx 80\%$ (\textit{b}, dashed-dotted red line) of the energy in the flow, respectively. Two-dimensional $xy$-planes are extracted at $z = L_z / 2$.} 
    \label{fig:pod}
\end{figure}

\subsection{Dataset description and training setup}
The data are cropped in the streamwise and jet-normal directions to reduce the time required for computing the POD modes since the original simulation was carried out in a much larger domain to avoid confinement effects. The cropped subdomain has dimensions $70h \times 30h \times 13.33h$. Furthermore, the data is uniformly downsampled in all spatial directions by a factor of two, with the resulting reduced dataset having dimensions $337 \times 146 \times 64$. 

POD generates the sequence of temporal modes, $\mathbf{\Sigma} \mathbf{V}^{\top}$, for the entire time series, that is split in two chunks that cover the intervals $\left[ 0 , t_{train} \right]$ and $\left[ t_{train} + 1 , K \right]$. We employed the first $t_{train} = 557$ snapshots for training the model, roughly $90 \%$ of the dataset, while the remaining $t_{test} = 80$ are used as test data. Data are spaced in time $\Delta t U / h = 2$, that is sufficient for resolving the dynamics of the large-scale coherent structures, totaling $23$ Gb in memory for the three components of the three-dimensional velocity field. 

The models are trained using the Adam algorithm \cite{Kingma2015ICLRADAM}. The learning rate is set variable via exponential decay, starting at $10^{-3}$ and decaying with a rate of $0.99$ over $1500$ epochs using batch size of $16$. The number of trainable parameters ranges from $8.9 \cdot 10^{5}$ in the smallest model to $7.5 \cdot 10^{6}$ in the largest one.

\section{Results}\label{sec:res}
We first show in fig.~\ref{fig:pod} the singular values and reconstruction from the POD of the turbulent viscoelastic jet. POD yields a finite set of optimally-ranked orthogonal modes. The most energetic modes, i.e., those with the highest eigenvalues, are related to the most dominant coherent structures in the flow. Therefore, the reconstruction of the data using a few dominant POD modes gives a low-dimensional representation of the flow based on the largest scales that is used later for training the model. This approach reduces the computational cost, where the number of degrees of freedom is significantly reduced---from hundred thousands of grid points to tens of modes. Panel \textit{a} shows the decay of the eigenvalues. Their magnitude decreases significantly until $N \approx 25$ modes, corresponding to $\approx 50 \%$ of the energy in the flow (see panel \textit{b}). The reconstruction using the first $25$ POD modes yields a representation of the flow given by the most energetic flow structures (see panel \textit{c}), where POD models the large-scale dynamics of the bulk flow, but it misses the smaller scale dynamics at the near-field and the wake of the jet. These can be partially recovered by increasing the number of POD modes; for instance, using the first $N = 125$ modes yields a reconstruction that contains $\approx 80 \%$ of the energy (see panels \textit{b} and \textit{d}), thus capturing less energetic dynamics and refining the model.

There is a clear drawback in this methodology: it is required using many modes for reconstructing faithfully turbulent flows, owing to the linear basis from POD. Nonlinear methods based on autoencoders achieve better compression rates \cite{Murata2020JFMConvAE}, though their basis are neither orthogonal nor ranked; hierarchical autoencoders rank their modes following their contributions to the reconstructed field, but modes remain non-unique \cite{Fukami2020PoFHierCNNAE}. Recent approaches based on $\beta$-variational autoencoders address this issue, whose near-orthogonal nonlinear basis provides a more interpretable reconstruction of turbulent flows using fewer modes compared to POD \cite{Eivazi2022ESWAbVAE}. However, purely machine learning approaches require big data, that is bigger if the flow is three-dimensional. On the other hand, POD is a nonparametric method, so the number of trainable parameters of the model is significantly reduced. Furthermore, the subspace from POD is physically interpretable---modes represent coherent structures in the flow---unlike for autoencoders, that is less physically interpretable and their subspace may contain many frequencies, some of them linked to chaotic dynamics, that make their training harder and the model less robust. Models that use the POD modes as input of the neural network constrain the solution to evolve in the subspace from POD, which is consistent with the temporal dynamics of the system, thus reducing the number of parameters and the amount training data \cite{AbadiaHeredia2025PoFPODDL}.

\begin{figure}[t!]
    \centering
    \includegraphics[scale=0.53, keepaspectratio]{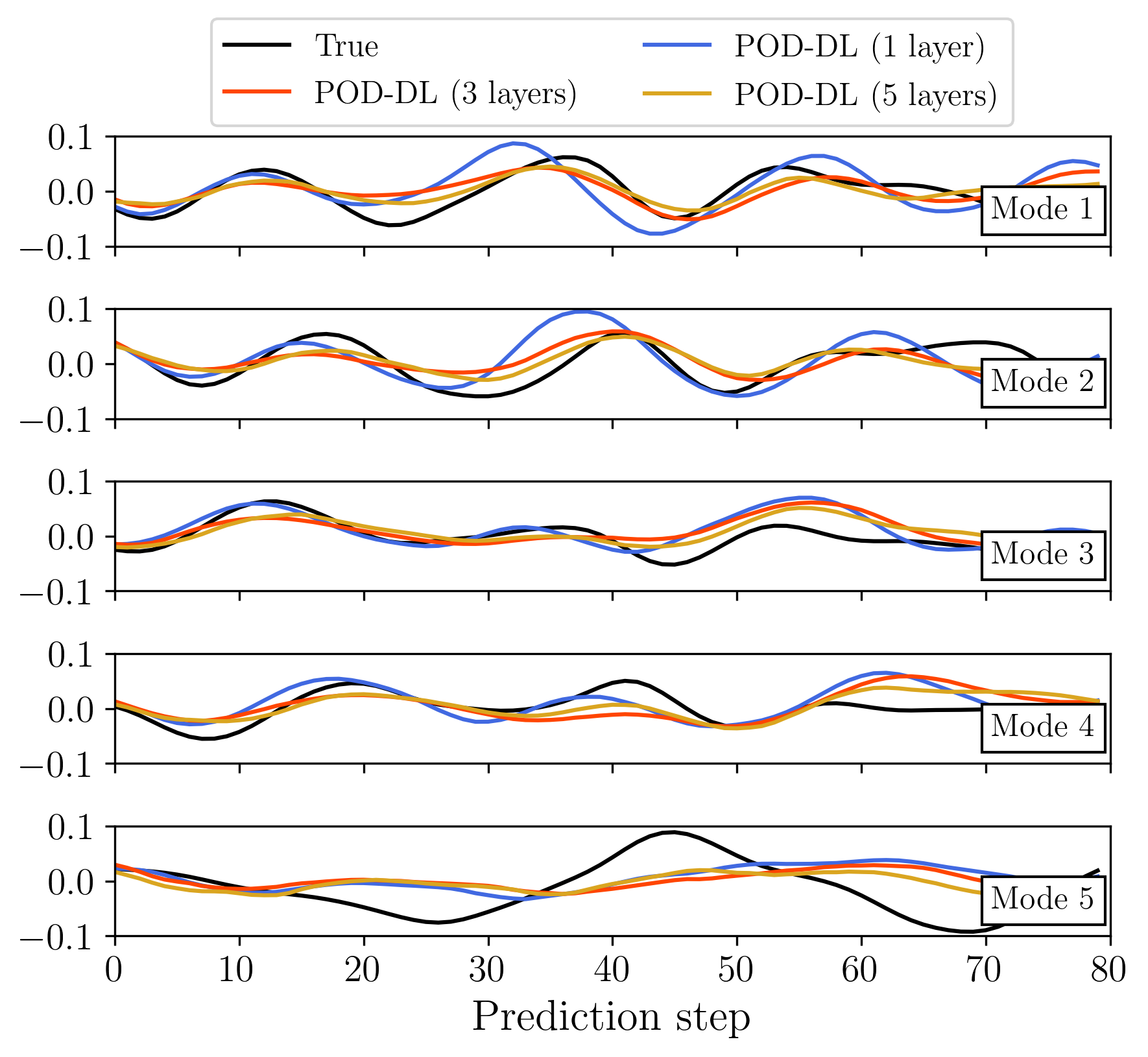}
    \put(-225,195){\small \textit{a)}}
    \includegraphics[scale=0.53, keepaspectratio]{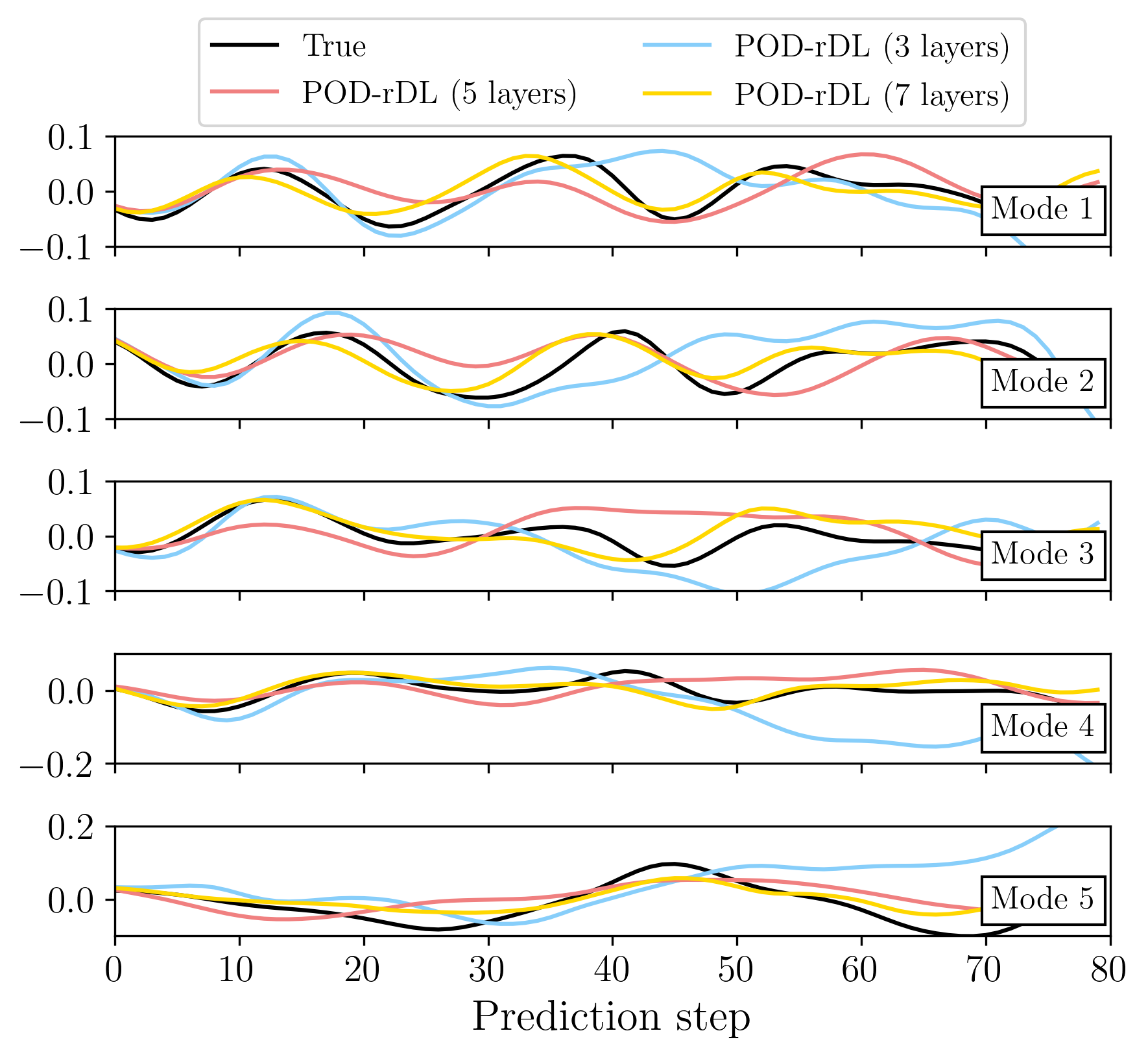}
    \put(-225,195){\small \textit{b)}}
    \caption{\textbf{Prediction in the space of POD modes.} Trajectory of the temporal coefficients and the predictions from the (\textit{a}) POD-DL and (\textit{b}) POD-rDL models.} 
    \label{fig:pred}
\end{figure}
\begin{figure}[t!]
    \centering
    \includegraphics[scale=0.6, keepaspectratio]{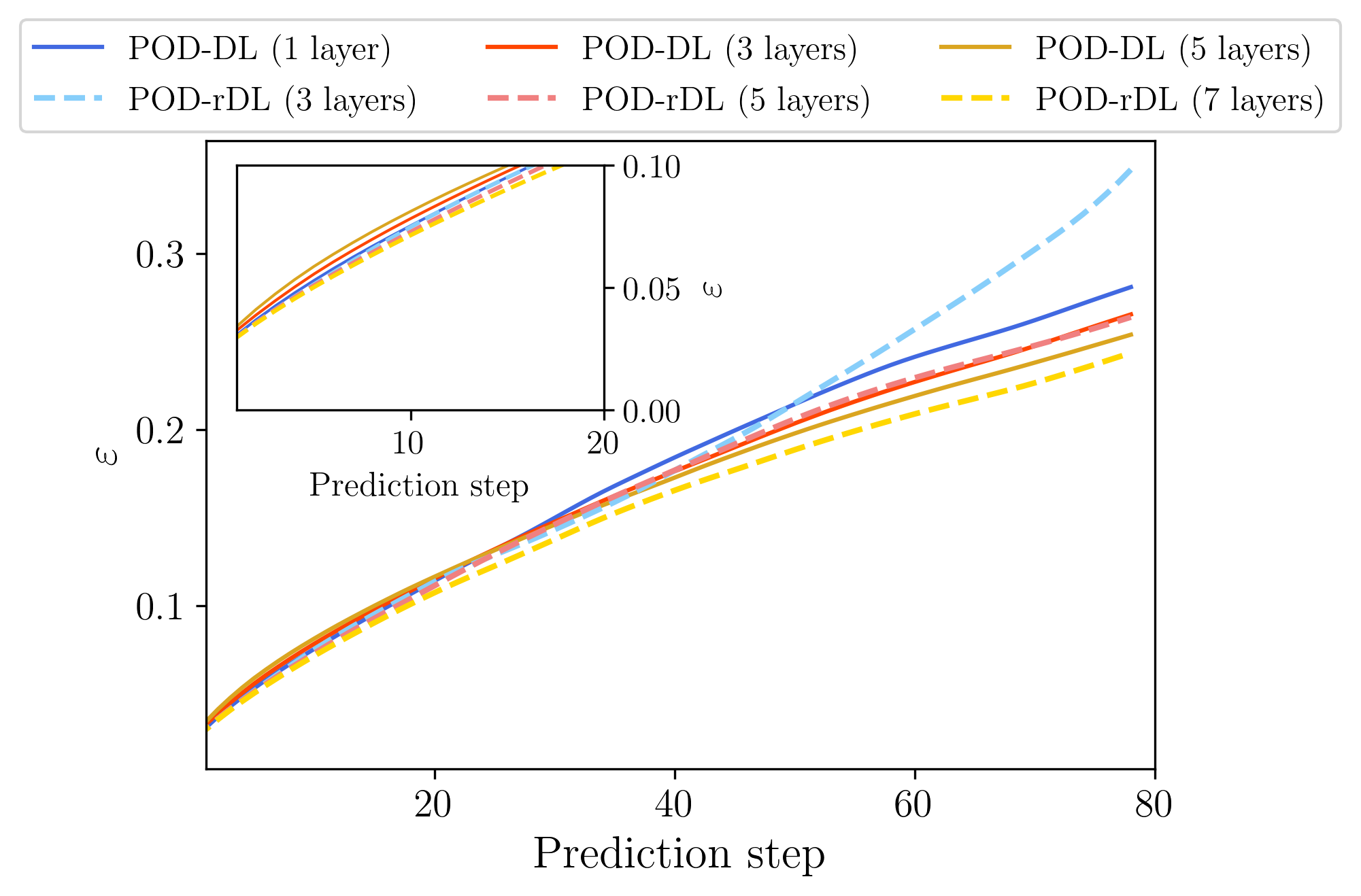}
    \put(-300,180){\small \textit{a)}}
    \put(-200,45){$\varepsilon (t) = \left\langle \left( \sum_{\tau = 0}^t \left( \mathbf{x}_{\tau} - \hat{\mathbf{x}}_{\tau} \right)^2 \right)^{1/2} \right\rangle$}
    
    \includegraphics[scale=0.38, keepaspectratio]{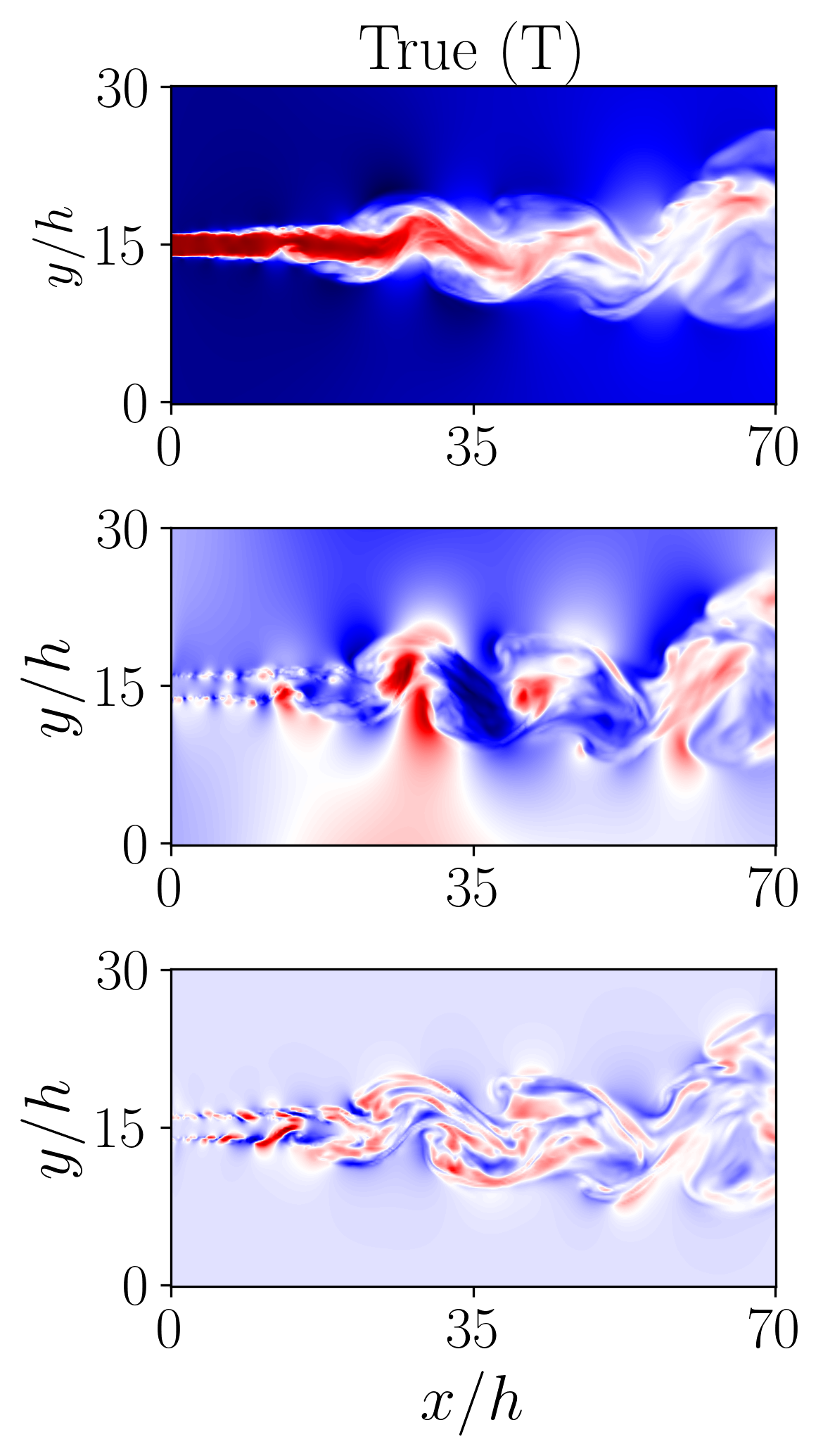}
    \put(-100,185){\small \textit{b)}}
    \includegraphics[scale=0.38, keepaspectratio, trim=30 0 0 00, clip]{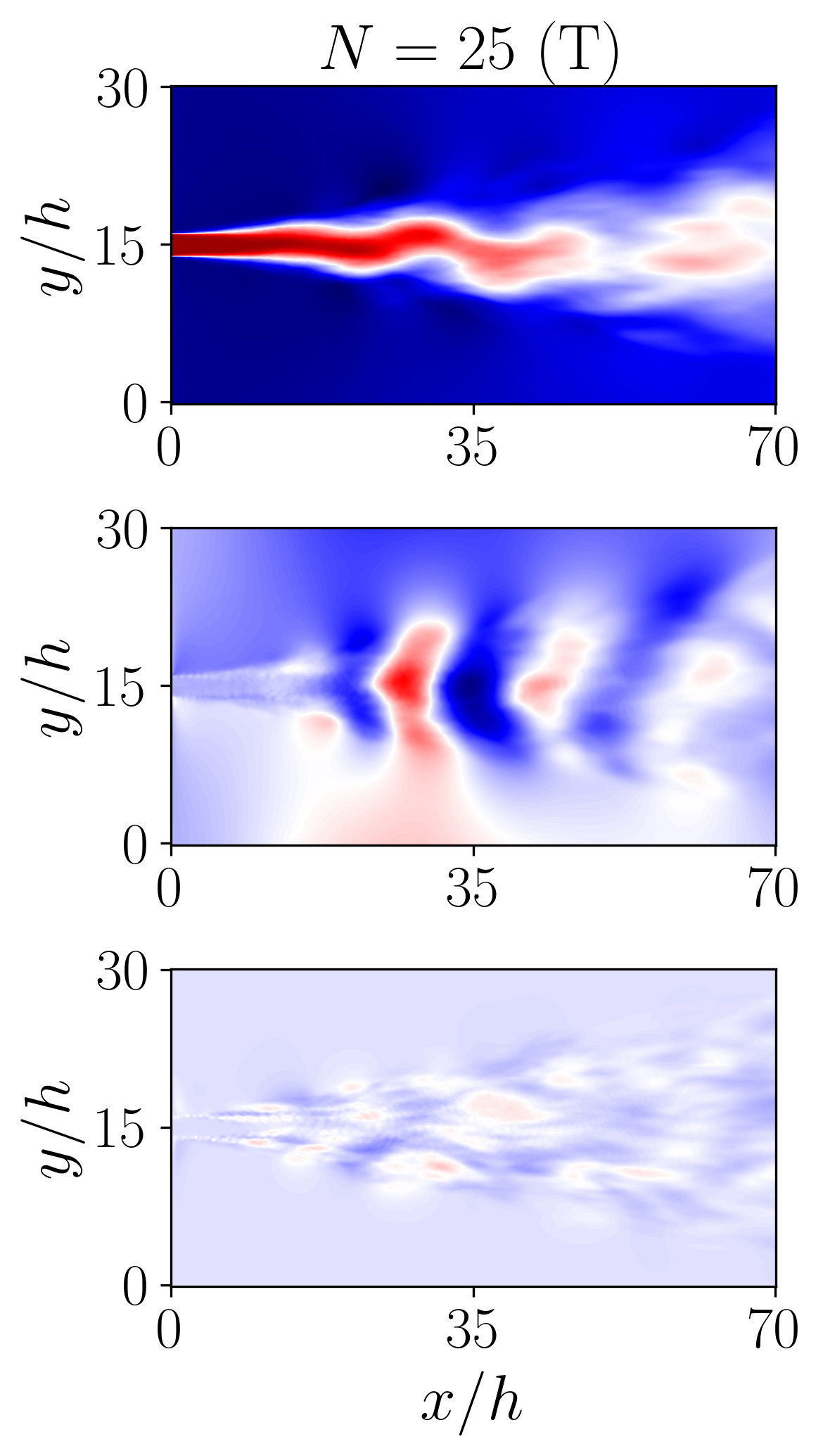}
    \put(-100,185){\small \textit{c)}}
    \includegraphics[scale=0.38, keepaspectratio, trim=30 0 0 0, clip]{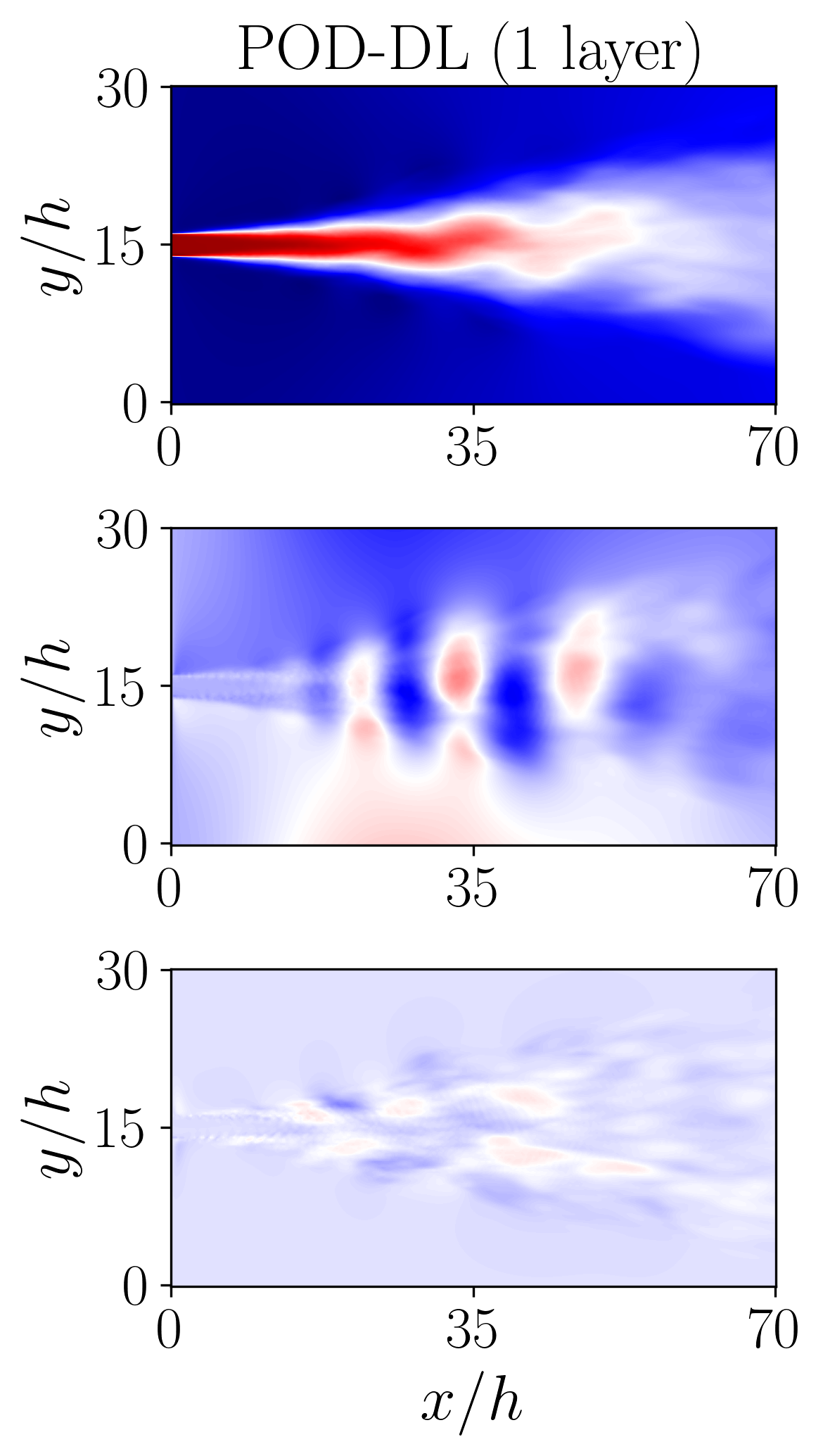}
    \put(-100,185){\small \textit{d)}}
    \includegraphics[scale=0.38, keepaspectratio, trim=30 0 0 0, clip]{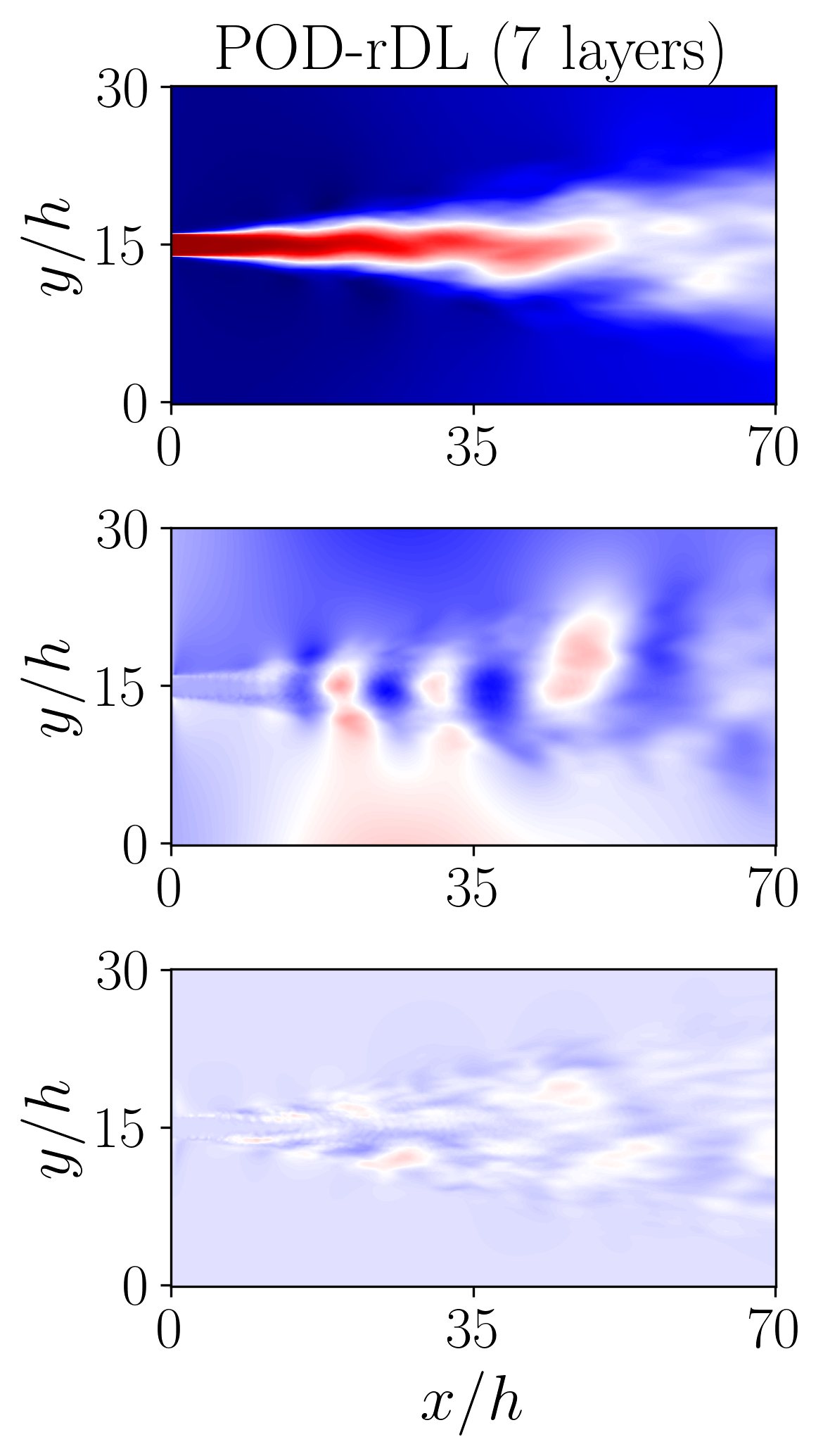}
    \put(-100,185){\small \textit{e)}}
    \includegraphics[scale=0.42, keepaspectratio, trim=240 9 0 0, clip]{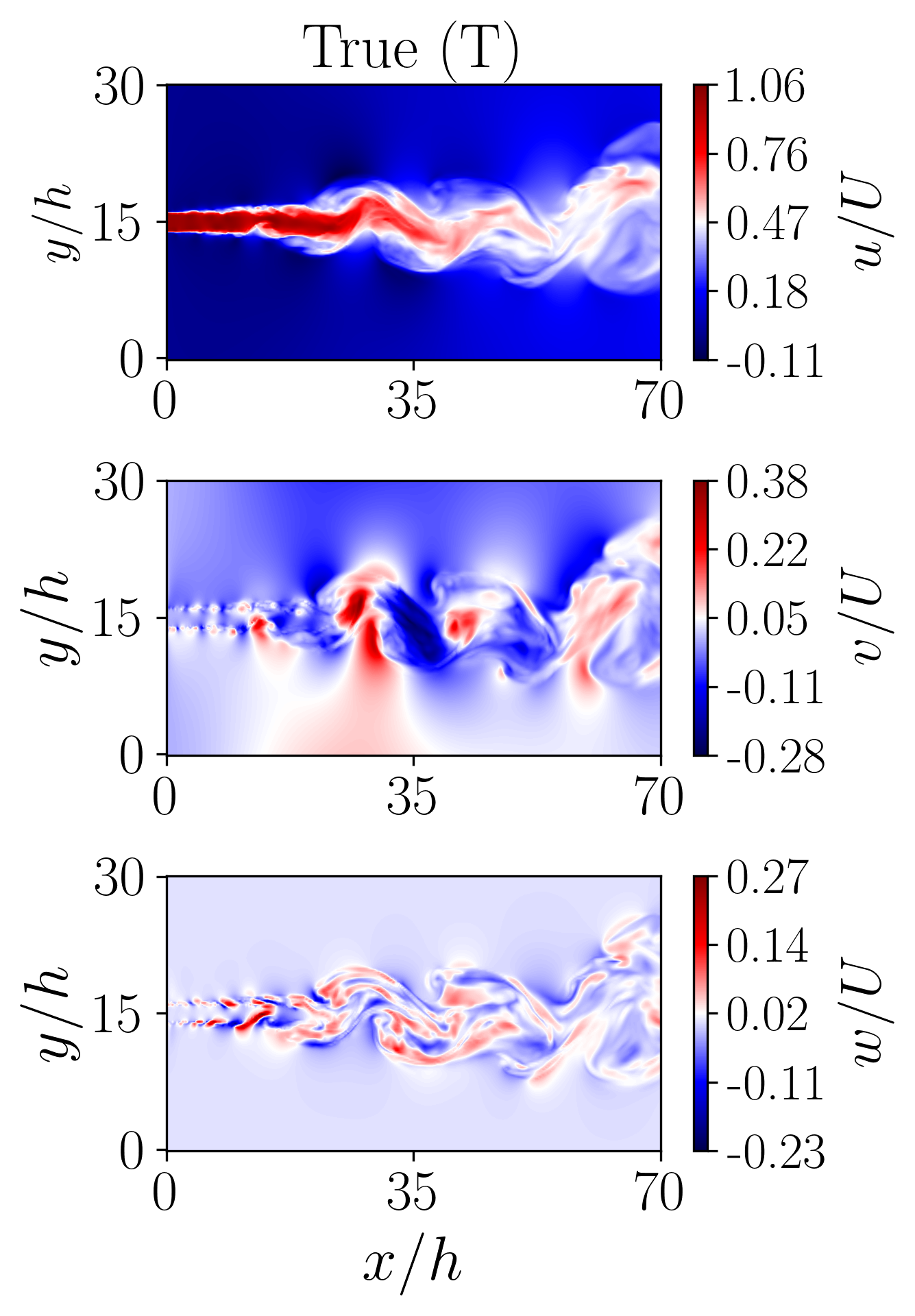}
    \caption{\textbf{Prediction of the velocity field.} Average prediction error over the temporal horizon (\textit{a}). Reconstruction of the streamwise (upper row), jet-normal (middle row) and spanwise (lower row) velocity field in the original space (\textit{c}-\textit{e}). The true data (\textit{b}) and their reconstruction using the first $25$ POD modes (\textit{c}) are compared to the prediction from the smallest (\textit{d}) and the largest (\textit{e}) models for the latest sample in the test data.} 
    \label{fig:error}
\end{figure}
To assess the performance of the models, that were trained using the first $25$ POD modes, we first show in fig.~\ref{fig:pred} the prediction of the temporal coefficients. We compare the true values of the first five modes with those from each model prediction. As expected for a chaotic system, the predictions diverge from the original trajectory, although they have similar quantitative performance than the true values. Moreover, the predictions from POD-DL seem to converge with layer depth, while those for the POD-rDL with $7$ layers (the deepest model) show a remarkable agreement with the reference values.

Next, we evaluate the accuracy of the predictions. Figure~\ref{fig:error}\textit{a} shows the error measured as the cumulative average of the L2 error norm to the temporal horizon, that is computed comparing the true velocity field, $\mathbf{x}$ and the reconstruction of the prediction in original space, $\hat{\mathbf{x}}$. Recall that the models generate predictions in an autoregressive way during inference, i.e., the output of the model is used in the next prediction step. As a consequence, errors accumulate as the temporal horizon gets longer, which could eventually lead the model to diverge. 

Overall, models are stable in time, except the POD-rDL with $3$ layers, that diverges after $40$ autoregressive generations. Regarding the POD-DL, shallow models perform better for short horizons, whereas deeper models improve their accuracy for longer horizons; conversely, deep POD-rDL models perform better at short and long horizons. This improvement is not surprising. As layer depth increases, models are able to represent more complex relationships from the data (more layers enable models to learn more abstract features). We observe that skip connections are the most effective strategy for training deeper models. The combination of skip connections with layer normalization enhances the generalizability of the model, owing to a smoother optimization of the neural network, \cite{He2016Residual, Ba2016arxivLayerNorm}, unlike the POD-DL, whose performance saturates sooner.

We now visualize the predictions in the original space. The velocity field is recovered from projecting the predicted temporal coefficients from POD back to physical space and adding the subtracted mean value. Figures~\ref{fig:error}\textit{d,e} show two-dimensional slices of the three-dimensional velocity field, that are compared to the true data (panel \textit{b}) and the reconstruction using $25$ POD modes (panel \textit{c}). We choose in this comparison the shallowest (the smallest POD-DL, panel \textit{d}) and the deepest (the largest POD-rDL, panel \textit{e}) models. Recall the models are trained to match the POD reconstruction rather than the original data. Therefore, models trained with few POD modes interpretate the flow dynamics based on the largest, or the most energetic, structures, neglecting the smaller ones (POD already underestimates the jet-normal $v$ and spanwise $w$ velocity fields). After $80$ autoregressive steps, both predictions have a good agreement with the reconstruction from POD, though the velocity magnitude remains somewhat closer to the true one (panel \textit{c}) in the POD-rDL, particularly in the far field, where the POD-DL underestimates the value of the streamwise velocity $u$.

\begin{figure}[t!]
    \centering
    \includegraphics[scale=0.65, keepaspectratio, trim=0 30 60 0, clip]{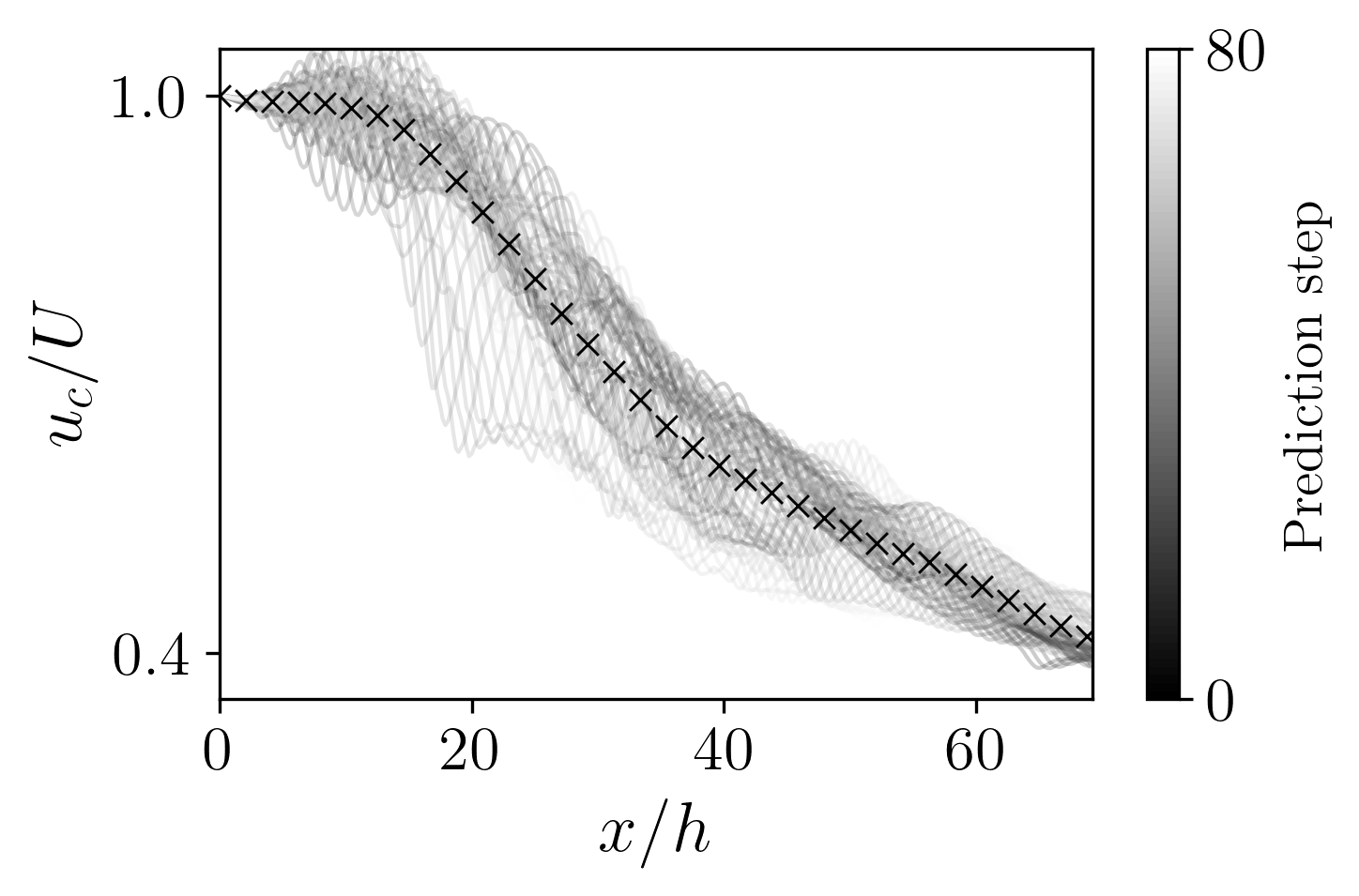}
    \includegraphics[scale=0.65, keepaspectratio, trim=0 30 0 0, clip]{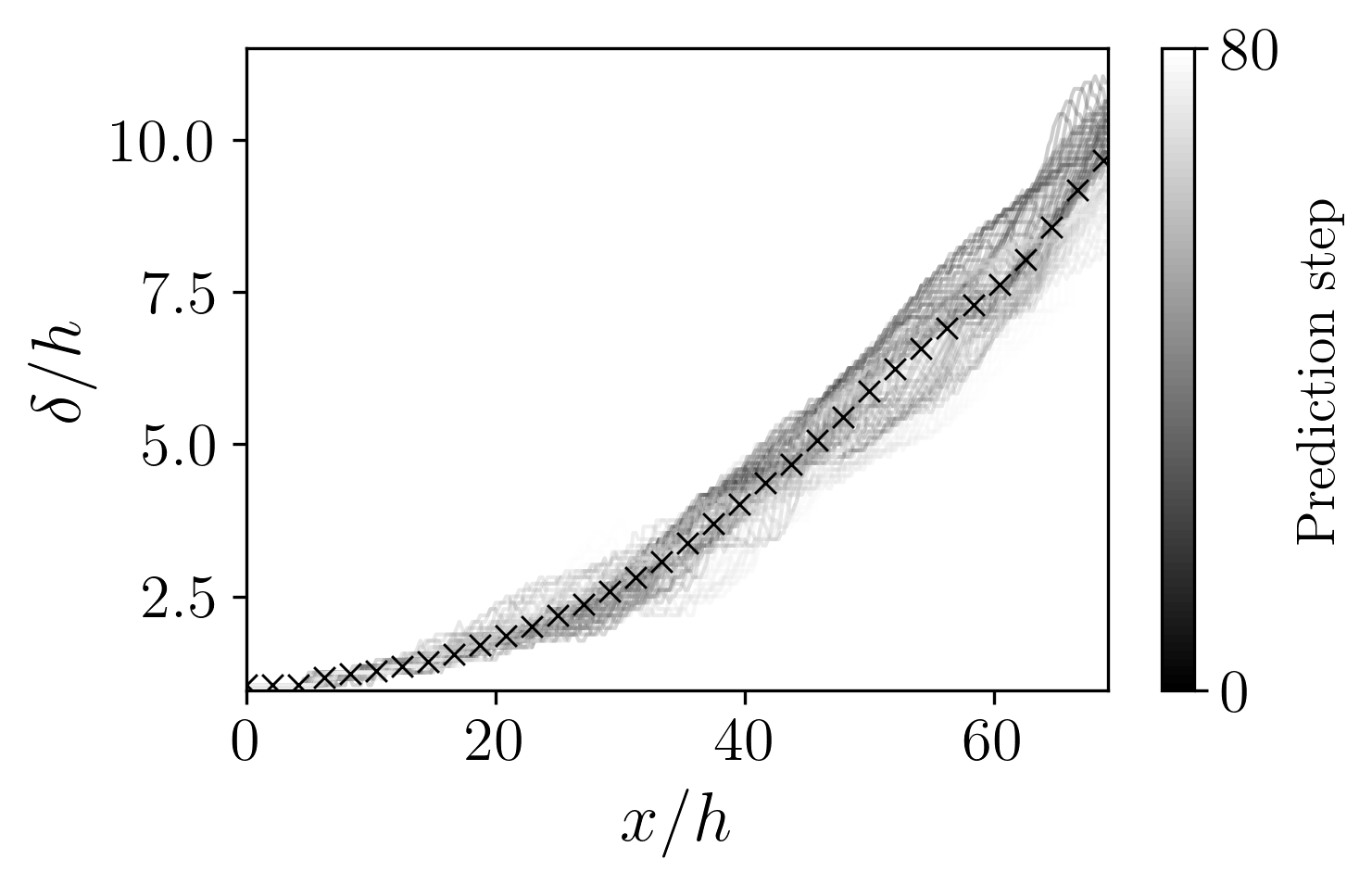}
    \put(-405,125){\small \textit{a)}}
    \put(-210,125){\small \textit{b)}}

    \includegraphics[scale=0.65, keepaspectratio, trim=0 30 60 0, clip]{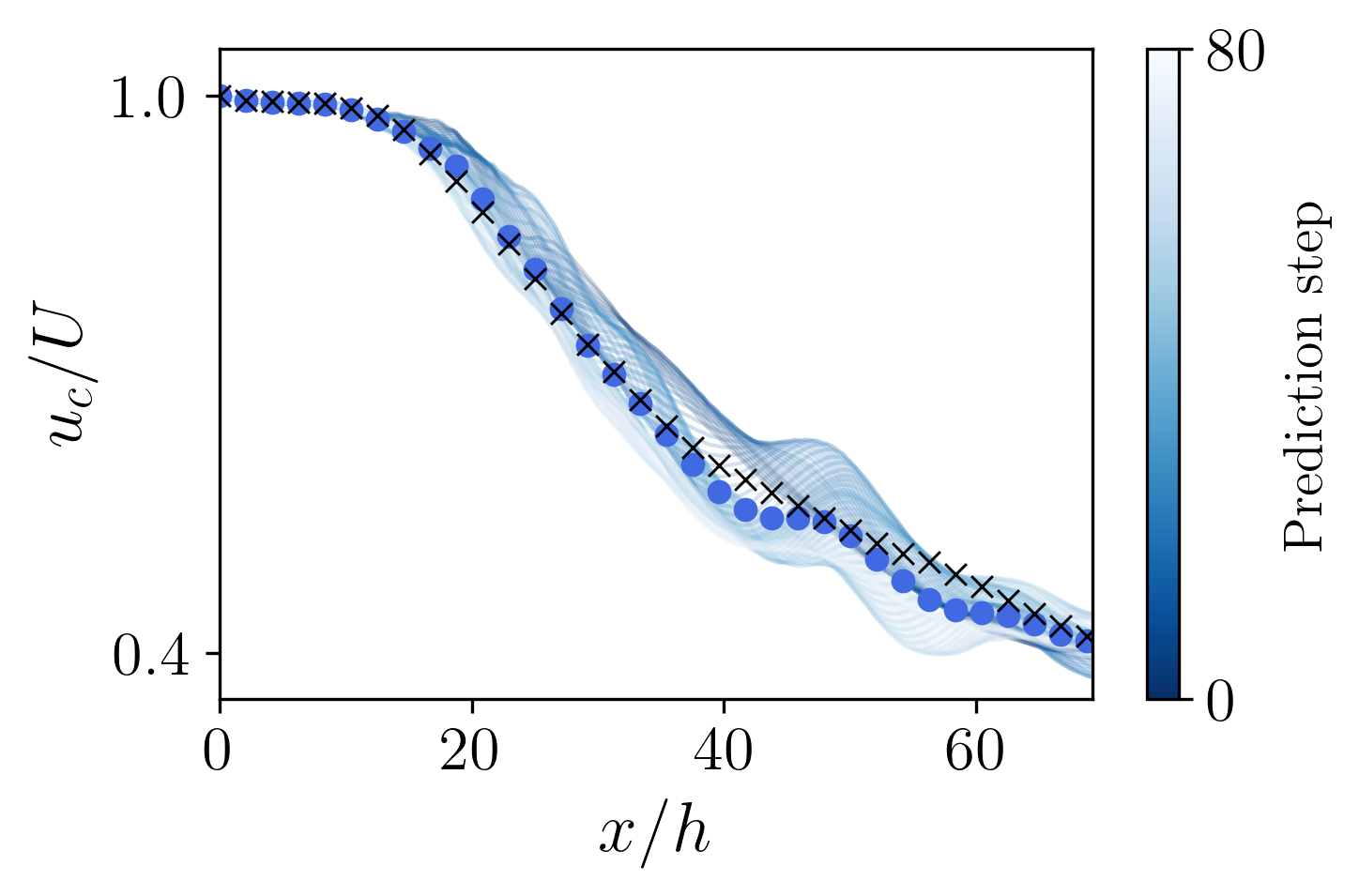}
    \includegraphics[scale=0.65, keepaspectratio, trim=0 30 0 0, clip]{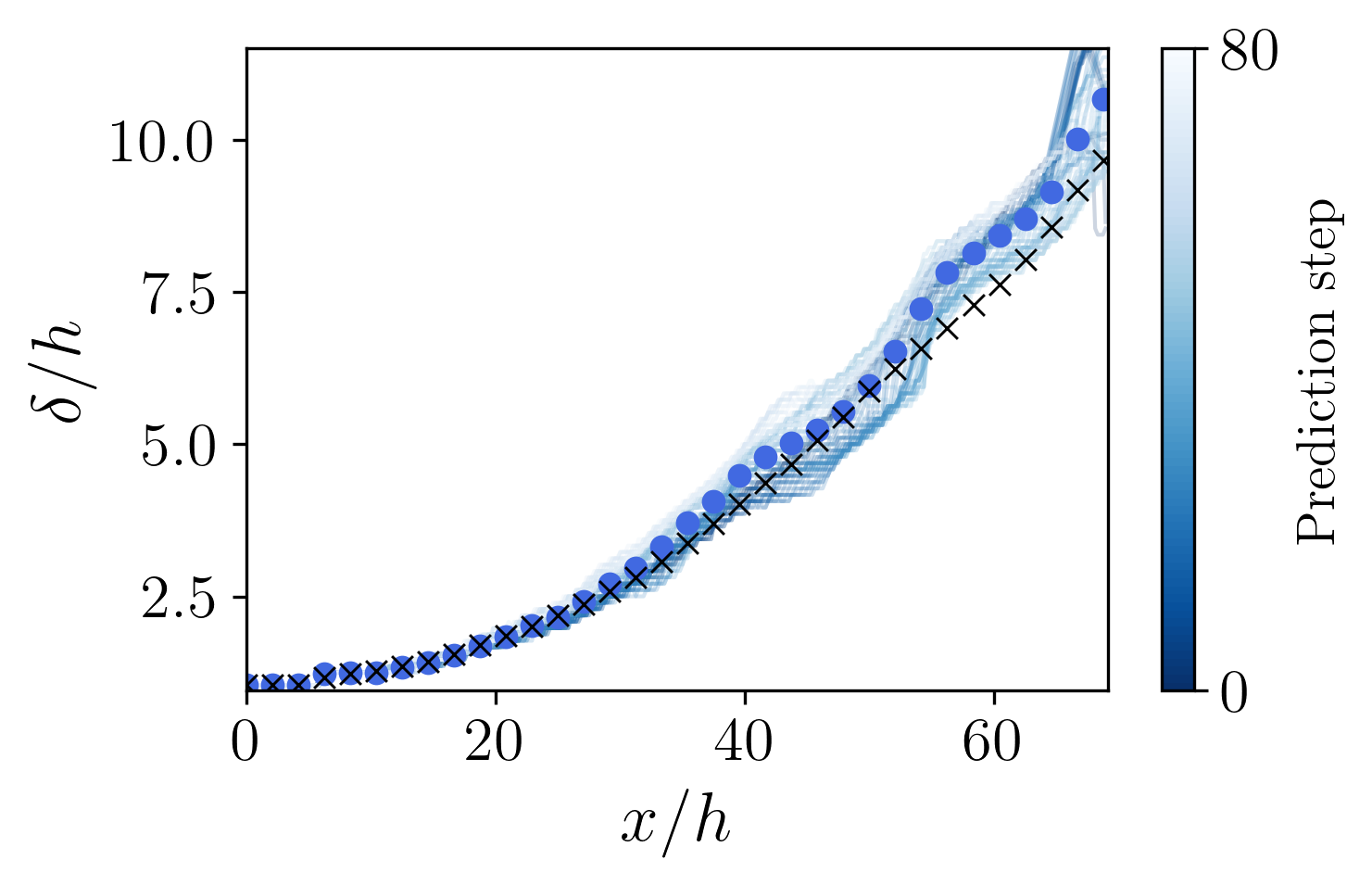}
    \put(-405,125){\small \textit{c)}}
    \put(-210,125){\small \textit{d)}}

    \includegraphics[scale=0.65, keepaspectratio, trim=0 0 60 0, clip]{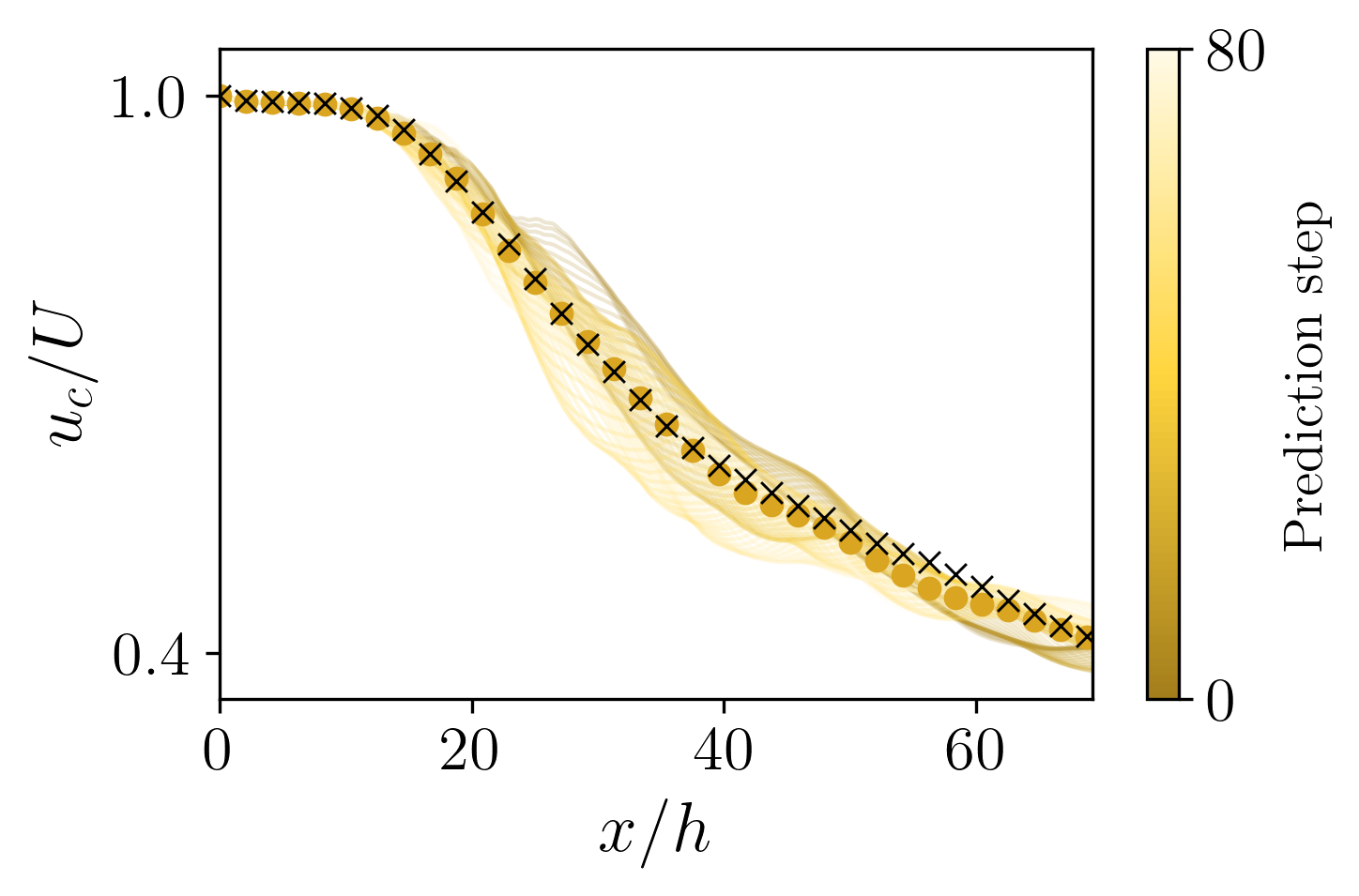}
    \includegraphics[scale=0.65, keepaspectratio, trim=0 0 0 0, clip]{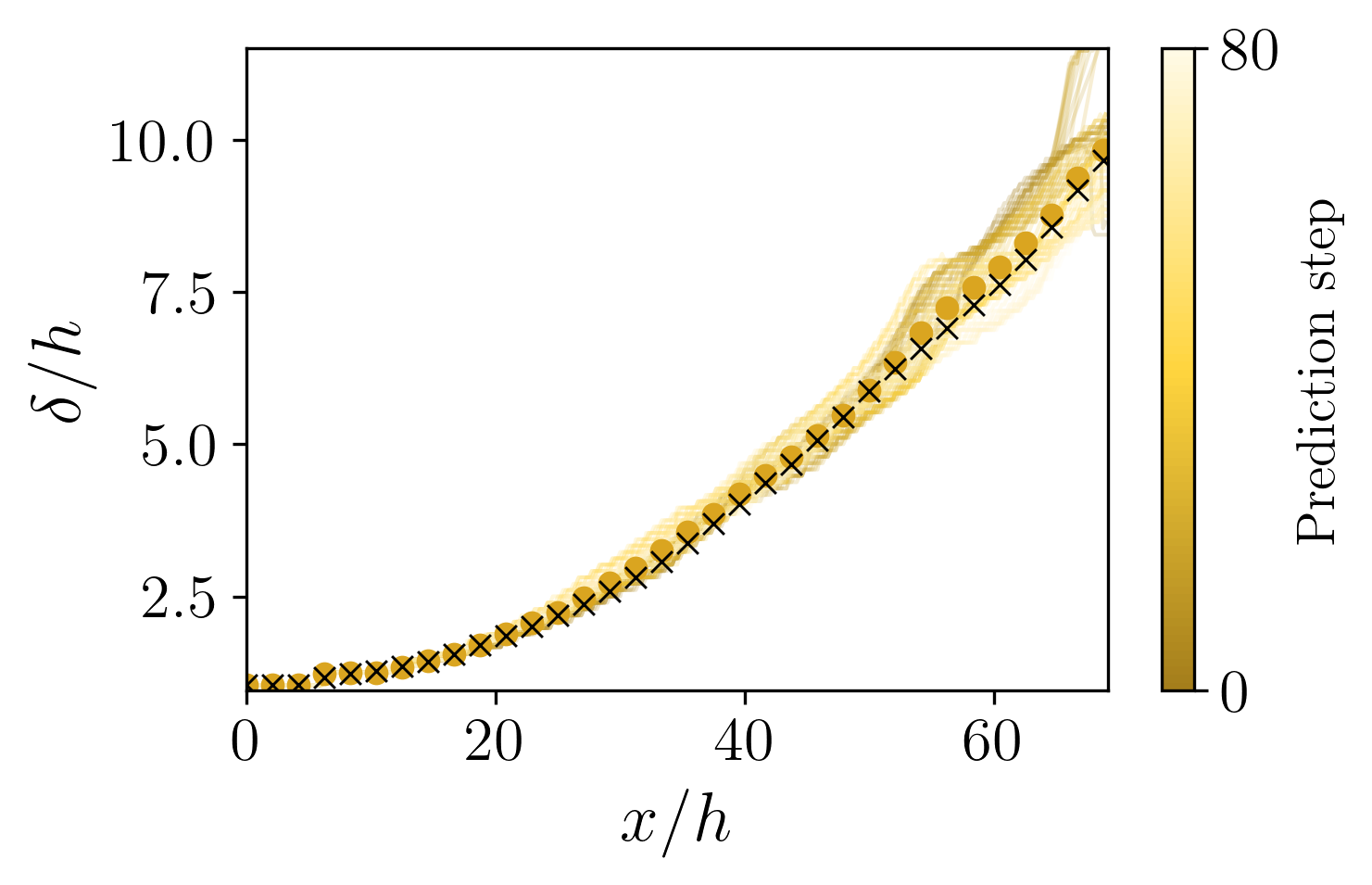}
    \put(-405,145){\small \textit{e)}}
    \put(-210,145){\small \textit{f)}}
    \caption{\textbf{Centerline velocity (a, c, e) and jet thickness (b, d, f).} True values (a, b) are compared to the prediction from the smallest POD-DL (c, d) and the largest POD-rDL (e, f) models. Solid lines indicate average in spanwise, and markers in spanwise and time.}
    \label{fig:vcdelta}
\end{figure}

We further assess the quality of the predictions by computing relevant statistics of the jet, namely the centerline velocity and jet thickness, in fig.~\ref{fig:vcdelta}. The true values (no decomposition, panels \textit{a} and \textit{b}) are compared to the predictions from the POD-DL (panels \textit{c} and \textit{d}) and the POD-rDL (panels \textit{e} and \textit{f}). Overall, both POD-DL and POD-rDL represent well the long-term behavior of both centerline velocity and jet thickness up to $x \approx 40h$, though the predictions from POD-rDL remain closer to the true values, especially in the wake, where POD-DL experiences larger deviations. However, the average quantities are not dominated by the mean contribution, as indicated by the instantaneous predictions in solid line, where the models are able to improve the prediction compared to just using the mean, although not fitting the true instantaneous values since being trained over the low-dimensional representation from POD.

\begin{figure}[t!]
    \centering
    \includegraphics[scale=0.5, keepaspectratio]{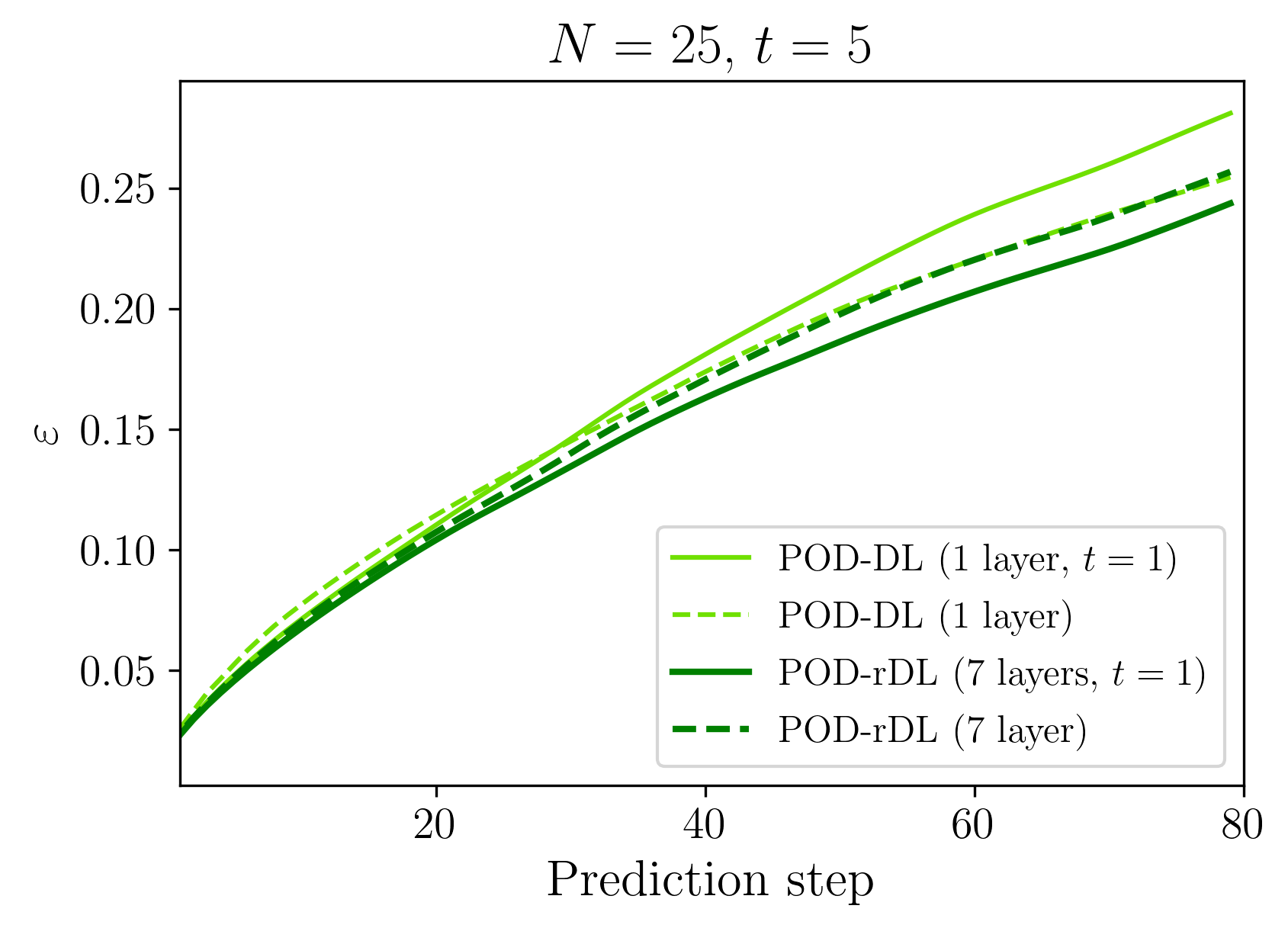}
    \put(-195,140){\small \textit{a)}}
    \includegraphics[scale=0.5, keepaspectratio]{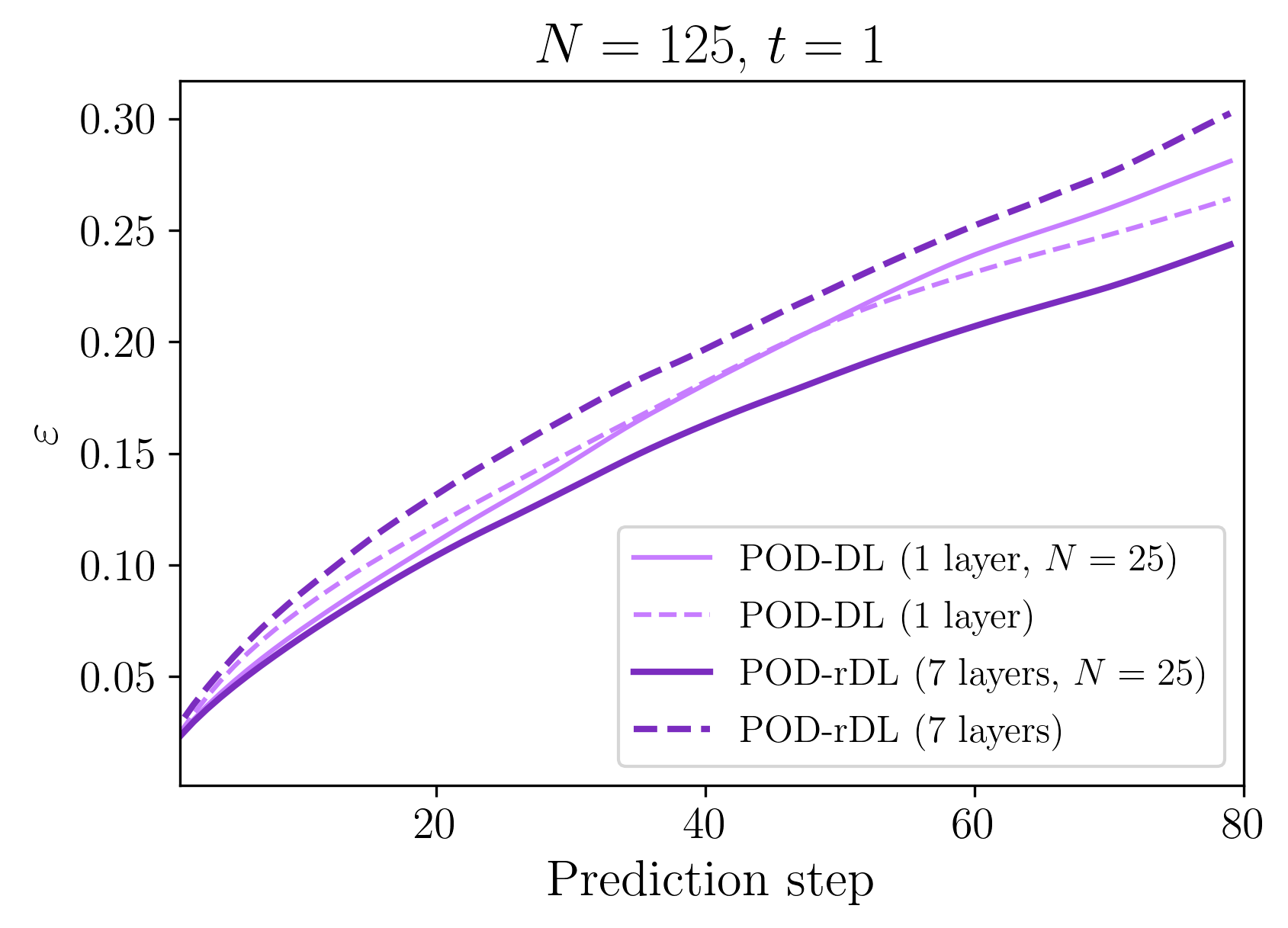}
    \put(-195,140){\small \textit{b)}}
    
    \includegraphics[scale=0.5, keepaspectratio]{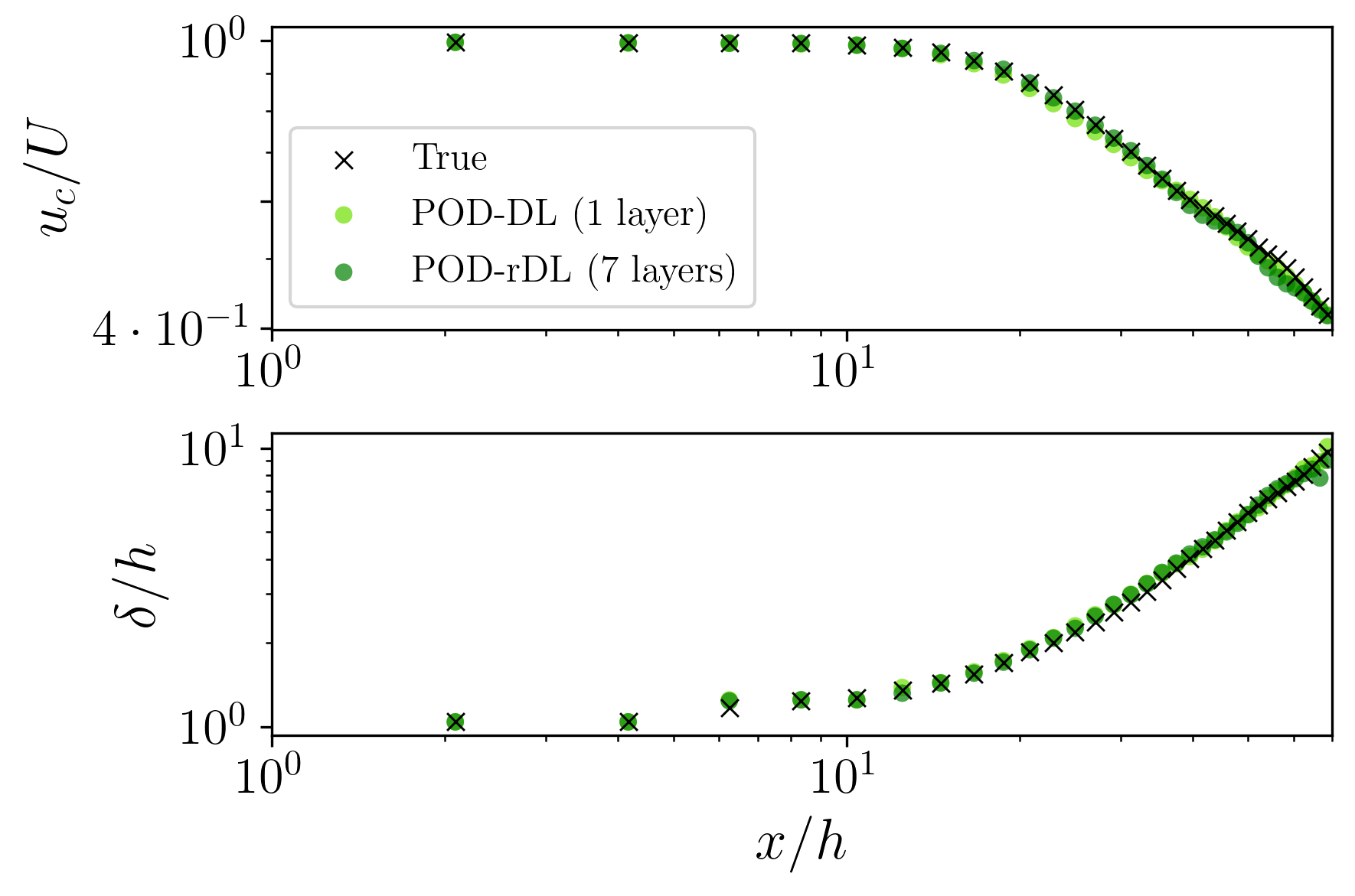}
    \put(-195,145){\small \textit{c)}}
    \includegraphics[scale=0.5, keepaspectratio]{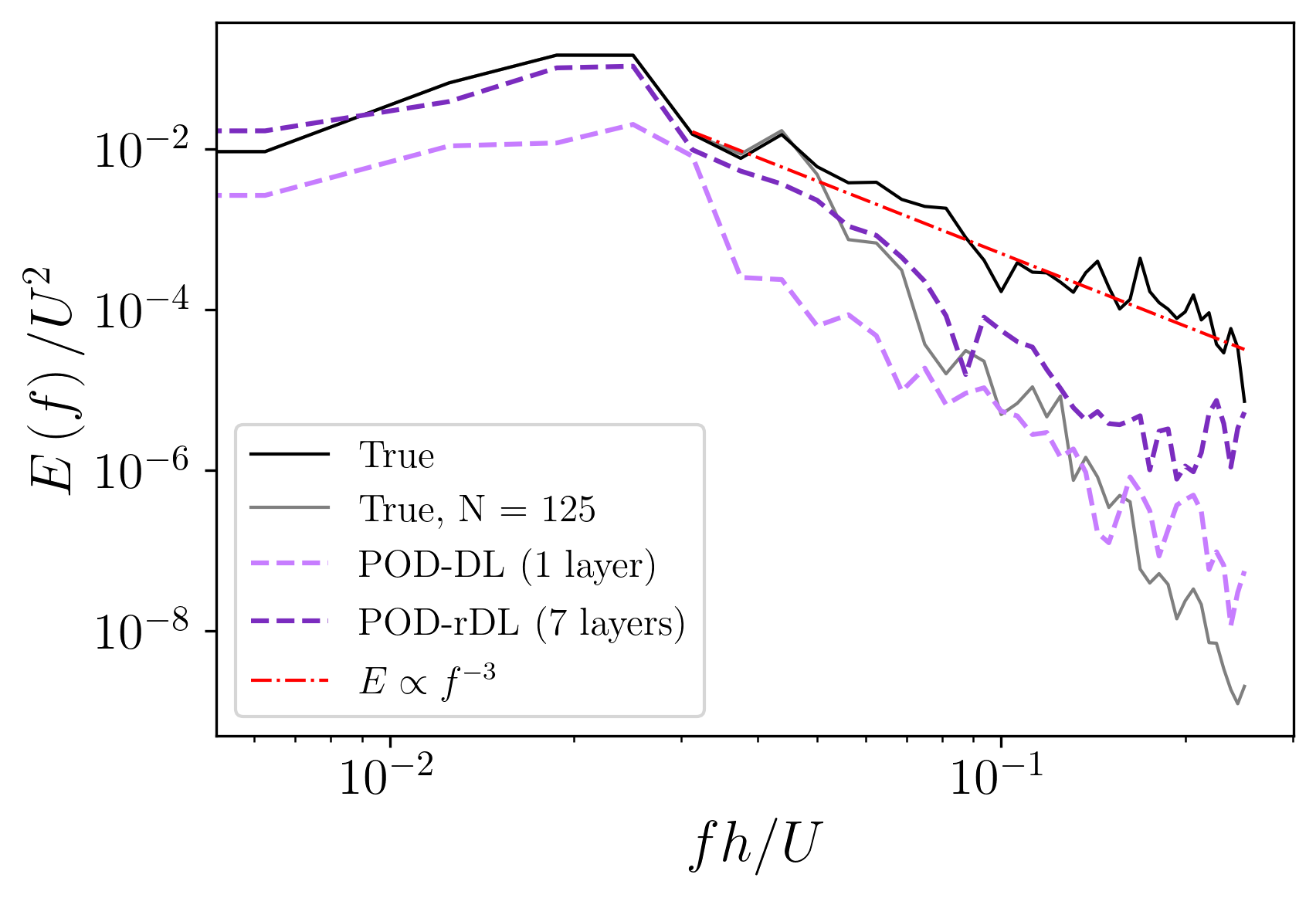}
    \put(-195,145){\small \textit{d)}}

    \includegraphics[width=\textwidth, keepaspectratio, trim=0 30 0 0, clip]{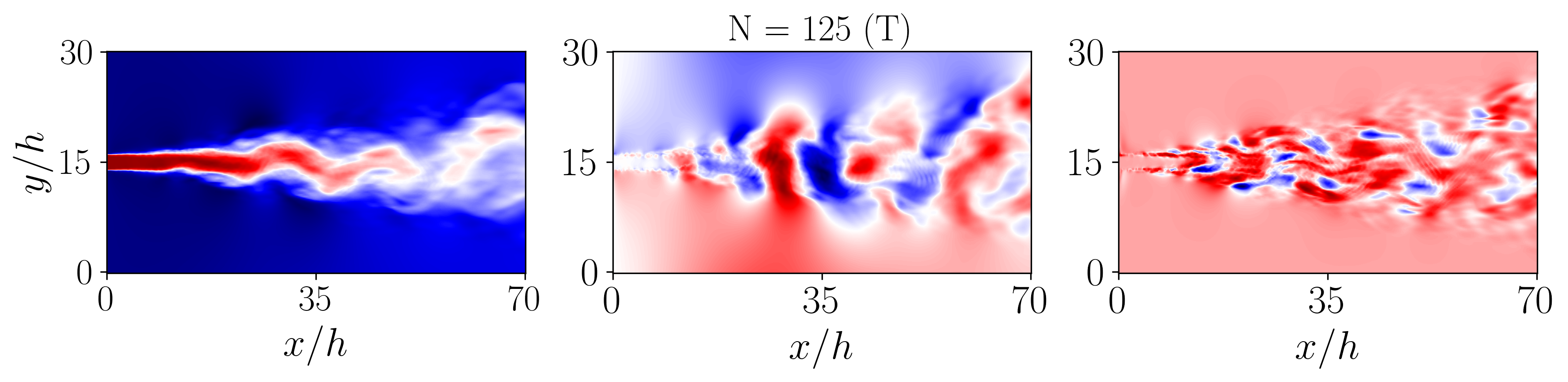}
    \put(-450,90){\small \textit{e)}}

    \includegraphics[width=\textwidth, keepaspectratio, trim=0 0 0 0, clip]{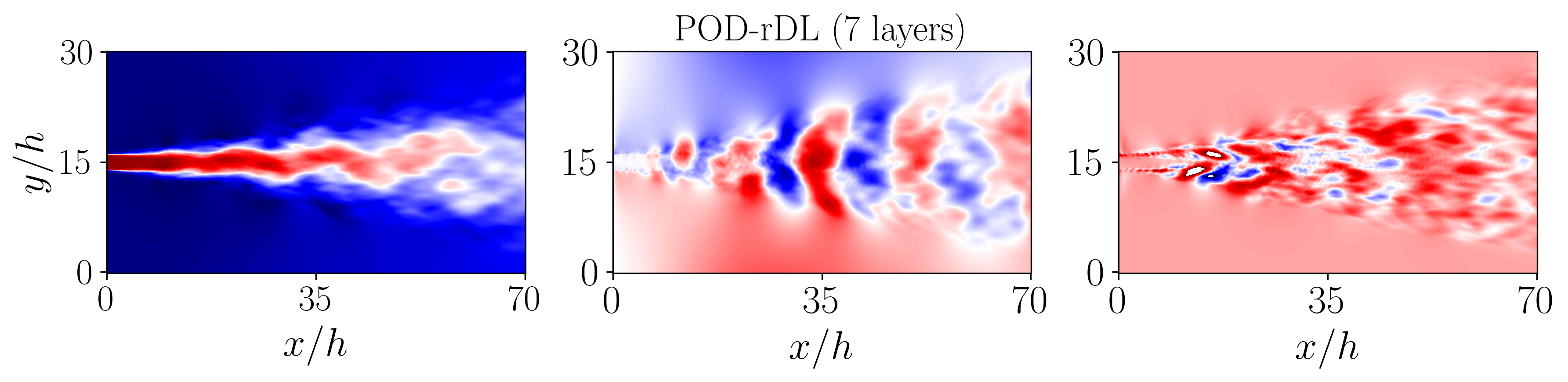}

    \hspace{+20pt}
    \includegraphics[scale=0.6, keepaspectratio]{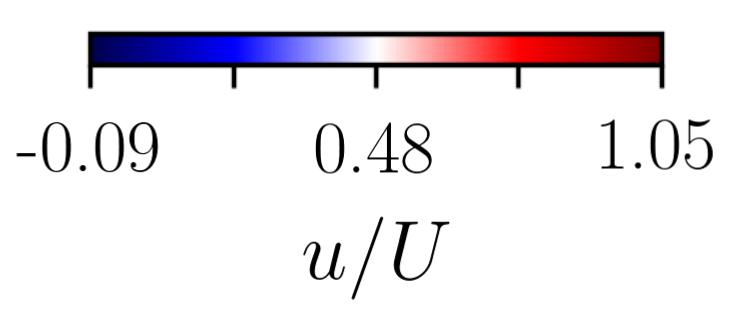}
    \hspace{+30pt}
    \includegraphics[scale=0.6, keepaspectratio]{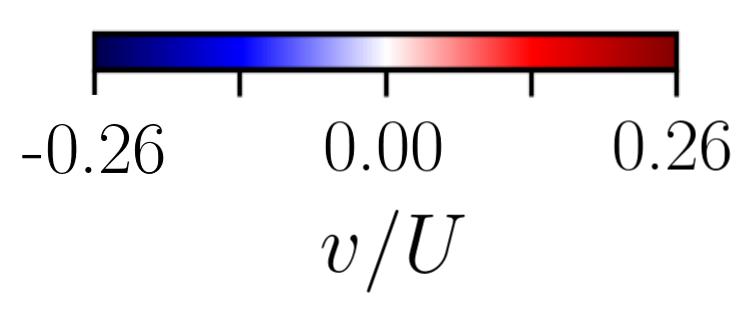}
    \hspace{+30pt}
    \includegraphics[scale=0.6, keepaspectratio]{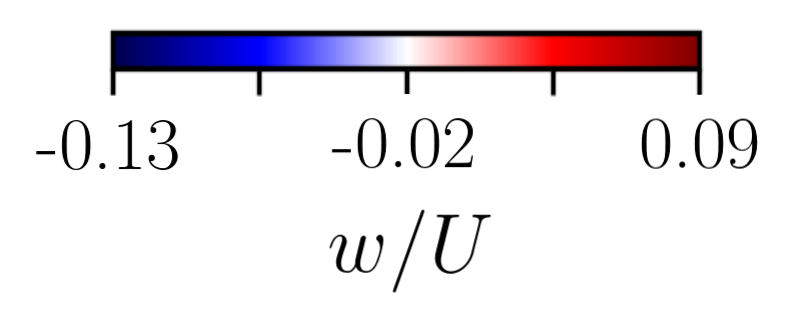}
    
    \caption{\textbf{Analysis of the predictions for larger prediction step $t$ and number of POD modes $N$.} Average prediction error over the temporal horizon for (\textit{a}) $t = 5$ and (\textit{b}) $N = 125$. The predictions are further evaluated by computing the centerline velocity $u_c$ and jet thickness $\delta$ (\textit{c}), the energy spectra (\textit{d}), and the three-dimensional velocity field (\textit{e}). Colors match between panels, whereas the velocity field is reconstructed for the POD-rDL model with $N = 125$.} 
    \label{fig:nmodessteps}
\end{figure}
We have shown so far the performance of the models for predicting the velocity field one step at a time using a few POD modes related to the most energetic structures in the flow. We conclude exploring briefly whether the models are also able to work with richer input spaces (adding more POD modes to the input) or longer autoregressive steps. We summarize our findings in fig.~\ref{fig:nmodessteps}. The growth of the prediction error remains somewhat similar if we increase the prediction step $t$ (panel \textit{a}) or the number of POD modes $N$ (panel \textit{b}), though there is a slight improvement in the accuracy of POD-DL for $t = 5$, and a worsening of the predictions in the case of POD-rDL for $N = 125$. However, the error growth does not provide any information about whether the models learned properly the jet dynamics for long-term predictions. Therefore, we compute again relevant statistics to properly evaluate their performance.

First, the centerline velocity and jet thickness are compared for the case with longer prediction step (panel \textit{c}). Predictions are similar between POD-DL and POD-rDL, which are in good agreement with the reference values. Nevertheless, the size of the models is significantly different: the POD-DL has roughly $\times 10$ fewer parameters compared to the POD-rDL. The good performance of POD-DL in this case indicates that layer depth is not necessary, but the construction of the model might have a greater influence. In particular, both models implement fully-connected layers after the LSTM block, although the way they work differs significantly. In the POD-DL, the MLP refines through nonlinear transformations the $t$ candidate steps mapped linearly from the output of the LSTM. On the contrary, the residual block in the POD-rDL learns a global representation first, which is refined through successive layers, before generating the following $t$ steps. The approach from POD-rDL is less expressive: after time pooling, the output of the residual block is mapped linearly to the output space. On the other hand, the MLP in the POD-DL model applies the nonlinearities per-step rather than to the full-context vector. 

However, the POD-rDL performs better in the case of more POD modes in the input space. We show in panel \textit{d} the energy spectra computed in a probe at $x = 40h$, that is within the fully-turbulent region of the jet (the scaling $-3$, reported in elasto-inertial and elastic turbulent jets \cite{yamani2021spectral, yamani2023spatiotemporal, soligo2023non}, fits for roughly a decade along the interval of inertial scales). It is also reported the spectrum computed at the same probe from the velocity field reconstructed using the first $125$ POD modes. It is evident that the POD-rDL has a better ability to represent the multiple time scale phenomena present in the data, where the energy spectrum is in better agreement with the true one. On the contrary, the POD-DL underestimates the energy content throughout all temporal scales of the spectrum (even those contained in the first $125$ modes), indicating that depth plays a relevant role on reproducing smaller scale dynamics if a large number of POD modes is considered. In this line, the reconstruction of the velocity field from the POD-rDL shows an excellent agreement with the true reconstruction (panel \textit{e}). Even though the model still underperforms reproducing the dynamics in the wake, the prediction is significantly improved in the near-field, where the POD-rDL is able to capture more complex dynamics in the flow compared to the model that uses $25$ POD modes.

\section{Conclusions}\label{sec:concl}
In this work, we used hybrid machine learning models for reduced-order modeling of a turbulent viscoelastic jet. In particular, we combined proper orthogonal decomposition with a deep network for making predictions in the space of POD modes. The methodology is based on the POD-DL \cite{AbadiaHeredia2022ESWAPODDL}, which combines POD with a neural network that implements recurrent and fully-connected layers. We also introduced an extension of the method named POD-DL with residual or POD-rDL, that uses skip connections for training deeper neural networks. The models are trained to predict the temporal coefficients from the POD, and their performance is assessed and compared predicting the full three-dimensional velocity field and relevant statistics from the jet, namely centerline velocity and jet thickness. It is observed that all models are able to reproduce the large-scale dynamics of the jet given a few POD modes as input, whereas deeper models, obtained from stacking LSTM layers or using skip connections, overall improve the accuracy of the predictions. In particular, the deepest POD-rDL model (with $7$ layers) is superior to the rest of models, and it is able of reproducing with the lowest reconstruction error the velocity field and jet statistics, specially at the wake, where the POD-DL model experiences the largest deviation with respect to the true data. Finally, we also explored the performance of the models by increasing either the length of the prediction step or the number of POD modes in the input space. We found that the POD-DL is able to generate predictions five steps at a time with an error comparable to the POD-rDL, but with a much smaller neural network, owing to its construction (each predicted step is processed nonlinearly, providing more expression power to the model if longer outputs are considered). On the other hand, layer depth ensures a better forecasting of large and smaller scales in the flow if a greater number of POD modes is considered, where the deepest POD-rDL model is able to reproduce more complex temporal dynamics.

\section*{Acknowledments}
The research was supported by the Okinawa Institute of Science and Technology Graduate University (OIST) with subsidy funding to M.E.R. from the Cabinet Office, Government of Japan. M.E.R. also acknowledges funding from the Japan Society for the Promotion of Science (JSPS), grant 24K17210 and 24K00810. C.A. and M.E.R. acknowledge the computer time provided by the Scientific Computing \& Data Analysis section of the Core Facilities at OIST, and by HPCI, under the Research Project grants hp250021 and hp250035. S.L.C. acknowledges the grant PID2023-147790OB-I00 funded by MCIU/AEI/10.13039/501100011033/FEDER, UE. This work has been partially done during the 2025 Madrid Turbulence Workshop, organized by Prof. J. Jim{é}nez and made possible by the European Research Council under the Caust grant ERC-AdG-101018287.

\bibliographystyle{unsrt}  

\end{document}